\documentclass[aps,twocolumn,groupedaddress,superscriptaddress]{revtex4-1}

\usepackage{graphicx,amsmath,amssymb}

\begin{document}

\begin{titlepage}
	
\title{Strain localisation above the yielding point in cyclically deformed glasses}
\author{Anshul D. S. Parmar} 
\affiliation{Theoretical Sciences Unit, Jawaharlal Nehru Centre for Advanced Scientific Research, Bengaluru, India.}
\affiliation{Tata Institute of Fundamental Research, Serilingampally Mandal, Ranga Reddy District, Hyderabad, India.}
\author{Saurabh Kumar}
\affiliation{Theoretical Sciences Unit, Jawaharlal Nehru Centre for Advanced Scientific Research, Bengaluru, India.}
\author{Srikanth Sastry}
\affiliation{Theoretical Sciences Unit, Jawaharlal Nehru Centre for Advanced Scientific Research, Bengaluru, India.}


\begin{abstract}
We study the yielding behaviour of a model glass under cyclic athermal quastistatic deformation computationally, and show that yielding is characterised by the discontinuous appearance of shear bands, whose width is about ten particle diameters  at their initiation, in which the strain gets localised. Strain localisation is accompanied by a corresponding change in the energies, and a decrease in the density in the shear band. We show that the glass remains well annealed outside the shear band whereas the energies correspond to the highest possible energy minima at the given density within the shear band. Diffusive motion of particles characterising the yielded state are confined to the shear bands, whose mean positions display movement over repeated cycles. Outside the shear band, particle motions are sub-diffusive but remain finite. Despite the discontinuous nature of their appearance, shear bands are {\it reversible} in the sense that a reduction in the amplitude of cyclic deformation to values below yielding leads to the healing and disappearance of the shear bands.
\end{abstract}

\maketitle
\end{titlepage}

The mechanical response of amorphous solids to applied stresses is of obvious importance in characterising their behaviour. Beyond the elastic regime at  small applied stresses, such response is characterised by plastic deformations, and beyond yielding, by flow. Considerations associated with yielding are of relevance to a wide range of phenomena, from irreversible deformation and failure in atomic and molecular glasses, such as metallic glasses, to the complex rheology of soft materials such as 
foams, emulsions, colloidal suspensions, and granular matter  \cite{Schuh2007a,Schall2010,Falk2010,Bonn2017c,Nicolas2017a}. Although yielding and flow may apparently be continuous and homogeneous for some yield stress fluids, it is a sharp, discontinuous event at the other end of the spectrum, as in brittle failure, characterised by localisation of strain and the formation of shear bands. Systems and questions of interest range from the mechanical properties of nanostructures to large scale phenomena such as mud-slides and earthquakes \cite{chester1998ultracataclasite,pekarskaya2001situ,cohen2006slip,Uhl2015b}.
Viewed as a nonequilibrium transition in a driven system, the phenomena associated with yielding have in recent years been investigated in a large number of studies experimentally, through computer simulations, and the analysis of {\it elastoplastic} and other models \cite{Hebraud1998,shi2005strain,Maloney2006,Dahmen2009,manning2009rate,Karmakar2010,Barrat2011,Dasgupta2012,Keim2013,Lin2014,Knowlton2014,Nagamanasa2014,denisov2015,regev2015reversibility,Liu2016,Itamar2016yielding,shrivastav2016yielding,leishangthem2017yielding,parisi2017shear,Urbani2017b,Jin,Ozawa2018a,Popovic2018a}. Many studies have focussed on the anisotropic interactions arising between localised plastic events or STZs \cite{argon1979plastic,falk1998dynamics,Picard2004,Puosi2014a,Dasgupta2012,Talamali2012c,Tyukodi2016c}, and,  in particular, how they may influence strain localisation.  Yielding has been analysed through the application of STZ theory \cite{Falk2010,manning2009rate}, described as a critical transition in analogy with depinning of manifolds in random media \cite{Dahmen2009,Lin2014}, a discontinuous transition associated with a spionodal \cite{Urbani2017b,parisi2017shear}, {\it etc.}, and the relationship between these descriptions is a subject of ongoing investigations ({\it e. g.},  \cite{Ozawa2018a,Popovic2018a}). 
The role of the degree of annealing in determining the nature of the yielding transition, and in the formation and character of shear bands, has increasingly been appreciated \cite{shi2005strain,leishangthem2017yielding,vasisht2017emergence,Ozawa2018a,Popovic2018a}. A particular situation in which the role of annealing becomes manifest is when an amorphous solid is subjected to oscillatory shear deformation
\cite{cohen2006slip,Fiocco2013,Fiocco2014,Knowlton2014,denisov2015,Radhakrishnan2016b,kawasaki2016macroscopic,leishangthem2017yielding,PRIEZJEV2018}. Under oscillatory or cyclic deformation of a glass, an increasing degree of annealing is observed as the amplitude of deformation is increased \cite{Fiocco2013,leishangthem2017yielding} (manifested by a decrease in energy) till the yielding strain is reached. Beyond the yielding strain, as described in detail below, the system yields through the formation of a shear band, within which the strain becomes largely localised. The overall energy of the glass increases from the yielding strain onwards. The width of the shear band increases with an increase in the applied strain amplitude, but at any given amplitude, reaches a steady state value for large numbers of cycles of deformation. 
However, since the system becomes inhomogeneous, one may inquire about the state of the glass within and outside the shear band. We investigate these changes in the present work, employing athermal quasistatic deformation of a model glass,  and show that the mechanical response of a cyclically deformed glass simultaneously displays features of aging or annealing outside the shear band, and of rejuvenation within. 



 We simulate the Kob-Andersen binary ($80:20$) mixture of $64000$ Lennard Jones
 particles at a reduced density of $\rho=1.2$.  Quadratic
 corrections are used to make the force and the potential energy continuous at
 the cutoff $r_{c~\alpha \beta} $(=$2.5\sigma_{\alpha \beta}$). The pair-wise interactions are defined as 

\begin{eqnarray}
U_{\alpha\beta}(r)&=&4 \epsilon_{\alpha\beta} \left[ \left( \frac{\sigma_{\alpha\beta}}{r} \right)^{12} - \left( \frac{\sigma_{\alpha\beta}}{r} \right)^{6} \right] \nonumber\\
& & + 4 \epsilon_{\alpha\beta}\left[c_{0 \alpha\beta} + c_{2 \alpha\beta}\left(\frac{r}{\sigma_{\alpha\beta}}\right)^{2}\right], r_{\alpha\beta} < r_{c\alpha\beta}, \nonumber \\ 
 &=& 0, \hspace{4.2cm}\text {otherwise}. \nonumber
\label{eqn:KABMLJmodel}
\end{eqnarray}
Here, indices $\alpha,\beta \in \{A,B\}$ refer to particle type,  $c_{0\alpha\beta}$ and $c_{2\alpha\beta}$ are chosen to ensure  
that the potential and its derivative at $r_{c\alpha\beta}$ vanish at the cutoff, 
and interaction parameters, defined with respect to the particles of type ``$A$", are:
$\epsilon_{AB}/\epsilon_{AA}=1.5$, $\epsilon_{BB}/\epsilon_{AA}=0.5$,
$\sigma_{AB}/\sigma_{AA}=0.80$ and $\sigma_{BB}/\sigma_{AA}=0.88$.
Energy and length are expressed throughout in units of $\epsilon_{AA}$ and
$\sigma_{AA}$, respectively.
Initial samples are generated by equilibrating the system at a high temperature $T=1$(in reduced units) using the Nos\'{e}-Hoover thermostat. 
We perform cyclic deformation for the most part over a range of amplitudes $\gamma_{max}$ above the yield strain amplitude ($\gamma_{y}\approx0.07$, \cite{Fiocco2013,leishangthem2017yielding}), from $0.07$ to $0.11$. We consider as starting configurations either 
the  energy minimum structures ({\it inherent structures}) obtained from minimising high temperature liquid configurations ($T=1$), or previously 
cyclically deformed  configurations at $\gamma_{max} = 0.07$ and $0.08$, which have reached steady state.  
The shear deformation is carried out using the athermal quasi-static (AQS) protocol,
 wherein each deformation step by a small strain increment is followed by energy minimisation using the conjugate gradient method. 
Samples are subjected to volume preserving shear along the $xz$ plane by incrementing
strain by small steps of $d\gamma_{xz}$ (here, $2\times 10^{-4}$) {\it via} the  coordinate transformation of $x^{\prime}=x+z~d\gamma_{xz},~y^{\prime}=y,~z^{\prime}=z$. The simulations are performed using  $LAMMPS$ \cite{Plimpton1995}.
Lees-Edwards periodic boundary conditions are employed in both the energy calculation
and  minimisation. We perform cyclic shear deformation ($0 \rightarrow \gamma_{max} \rightarrow 0 \rightarrow -\gamma_{max} \rightarrow 0$) repeatedly, till a steady state is achieved. As previously described, \cite{Fiocco2013,leishangthem2017yielding}, the number of cycles to reach steady state rise steeply as $\gamma_{y}$ is approached, but we do not discuss this feature further here.  For all the strain amplitudes above $\gamma_{y}$ studied, the steady state configurations display shear bands. 

To characterise the shear bands, we divide the configurations into slabs along the shear direction, and compute (a) the mean squared displacement ($\mathcal MSD(z)$) between particle positions in a slab centred at $z$ at strain $\gamma = 0$  separated by a full cycle of strain, and (b) the average energies of particles in each slab, $U(z)$, at the end of each cycle ({\it i. e.} we consider stroboscopic configurations). For a given cycle $i$, 
the $MSD$ for the $z^{th}$ slab is written as 
$MSD^{(i)}(z)=\sum_{j=1}^{n^{z}_{i}}({\bf r}^{z}_{i+1,j}-{\bf r}^{z}_{i,j})^{2}/n^{z}_{i}$,
 where $n_{i}^{z}$ and $\{{\bf r}^{z}_{i}\}$ represent the number of particles and 
 their positions in the $z^{th}$ slab of the stroboscopic configuration in the  $i^{th}$  cycle, respectively. 
 Similarly, the slab-wise energy can be defined as: 
 $U^{(i)}(z,i)=\sum_{j=1}^{n^{z}_{i}}u^{z}_{i,j} /2n_{i}^{z}$, where $u^{z}_{i,j}$ 
 is the interaction energy of the $j^{th}$ particle in the $z^{th}$ slab. The cycle index $i$ is not indicated unless necessary. 
 The profile of $MSD(z)$ is found to be well described by a Gaussian, 
$MSD(z) = MSD^{o} \exp \left (-(z -\langle z\rangle)^{2}/2\sigma^{2}\right)$, where $\langle z \rangle$ and $\sigma$ represents the mean position and width of the shear band, respectively. In order to characterise properties within and outside the shear band, we compute various partial (per particle) averages (other than slab-wise averages defined above), which we define here with the interaction energy as an example:  Partial averages are computed for the centre of the shear band ($U_{SB}(\sigma)$), most of the shear band (within $3 \sigma$) ($U_{SB}(3 \sigma)$), the rest of the system (outside $3\sigma$)  ($U_{rest}$), the slab of thickness $\sigma$ farthest from the centre of the shear band ($U^{'}(\sigma)$) and the global average value ($U$). In addition to the $MSD$ and energy $U$, we also compute the average displacement of particles per cycle, defined (for the full system) as  
 $\Delta r^{(i)} = \left.\sum_{j=1}^{N}|{\bf r}_{i+1,j} -{\bf r}_{i,j}| \right/ N$, for cycle $i$, and the mean squared displacement with respect to a reference configuration ${\bf r}^{0}_{j}$ for the $A$ particles as $\langle r^2_{A} \rangle ^{(i)} = \left.\sum_{j=1}^{N_A} ({\bf r}_{i,j} - {\bf r}^{0}_{j})^2 \right/ N_{A}$. 
   
 
\begin{figure*}[]
\centering
\includegraphics[scale=.23]{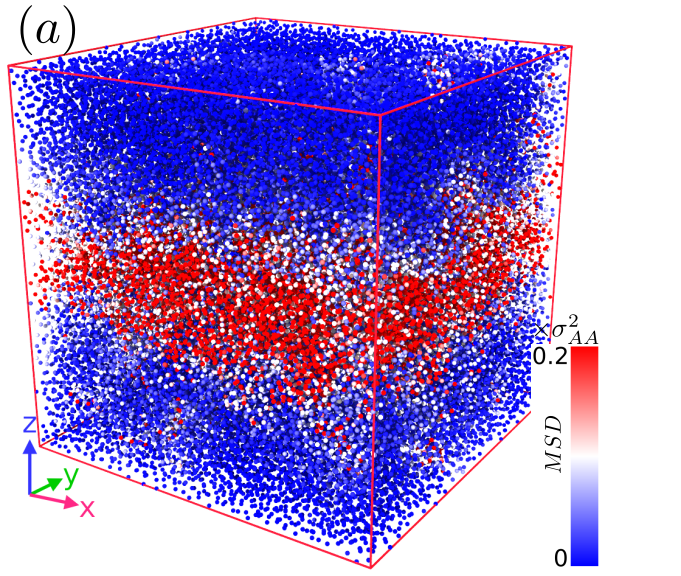} 
\includegraphics[scale=.35]{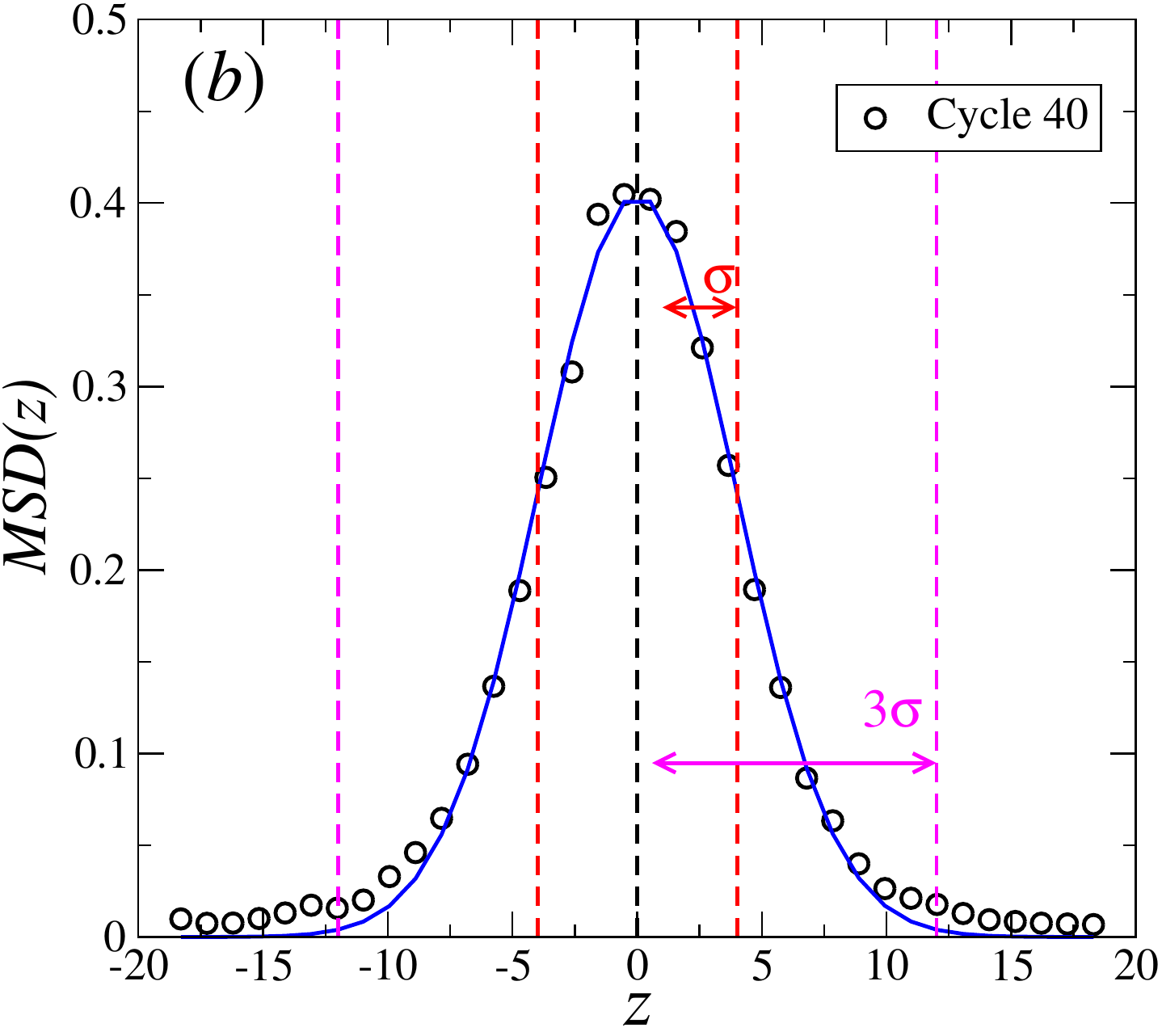} 
\includegraphics[scale=.35]{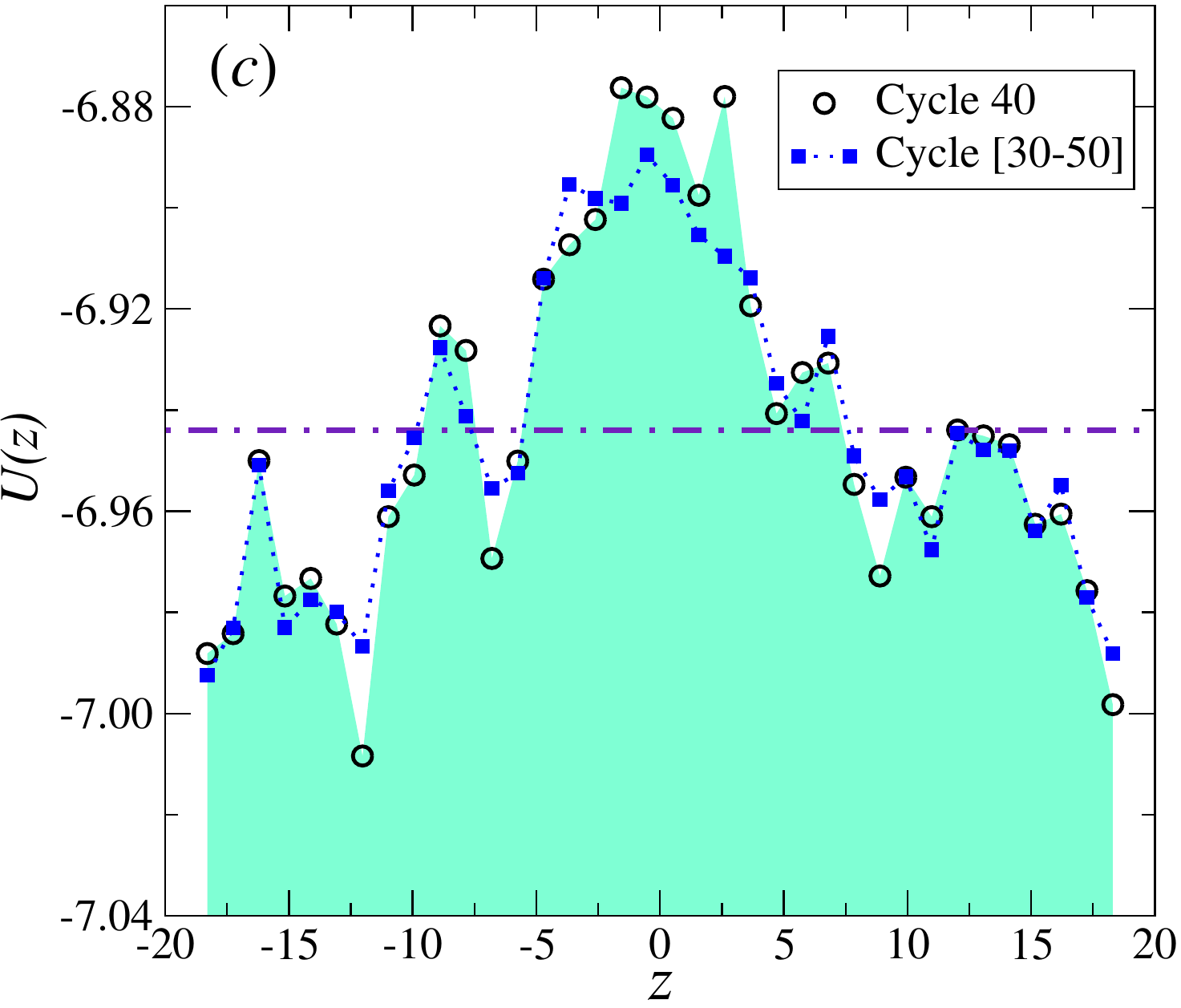} 
\caption{{(\emph a)} Snapshot of a configuration from the steady state for the strain amplitude $\gamma_{max}=0.09$, with the colour of the particle indicating the mean squared displacement $MSD$ over a strain cycle. Highly mobile particles ($MSD > 0.2~\sigma_{AA}^{2}$) are coloured in red, whereas particles in blue move considerably less. This snapshot shows that particle displacements are highly spatially correlated, forming a shear band. {(\emph b)}  $MSD(z)$ is shown as a function of the coordinate $z$ in the shear direction, along with a  Gaussian fit. 
For our analysis, we consider most mobile particles (within a width of $1\sigma$), and most of the particles (within $3\sigma$) in the shear band. {(\emph c)} The potential energy of mobile particles is seen to be higher than the mean potential energy represented by a horizontal line. The data presented by black circles corresponds to the $40^{th}$ cycle of strain, whereas data shown in blue boxes are averaged over $20$ cycles (from $30$ to $50$.}
\label{fig1}
\end{figure*}

In Fig. \ref{fig1}, we report the characterization of the shear band. We show
 configurations from the steady state for a given strain amplitude, $\gamma_{max}=0.09$. 
 The colour map is based on the mean squared displacement within a slab between 
 two consecutive strain cycles. Particles shown in red move more than 
 $0.2\sigma_{AA}^{2}$ ($MSD$ values roughly within $\langle z \rangle \pm \sigma)$). Fig. \ref{fig1} (b) 
 shows the $MSD(z)$ profile of the steady state configuration, which clearly shows 
 the existence of strain localisation or a shear band. The existence 
 of a shear band also gets reflected in the energy profile of  particles {\it vs.} the $z$ coordinate
 (see Fig. \ref{fig1}c); particles correspond to the band are likely to have higher 
 energy compared to the mean potential energy of the system.

\begin{figure*}[htp]
\centering
\includegraphics[scale=.45]{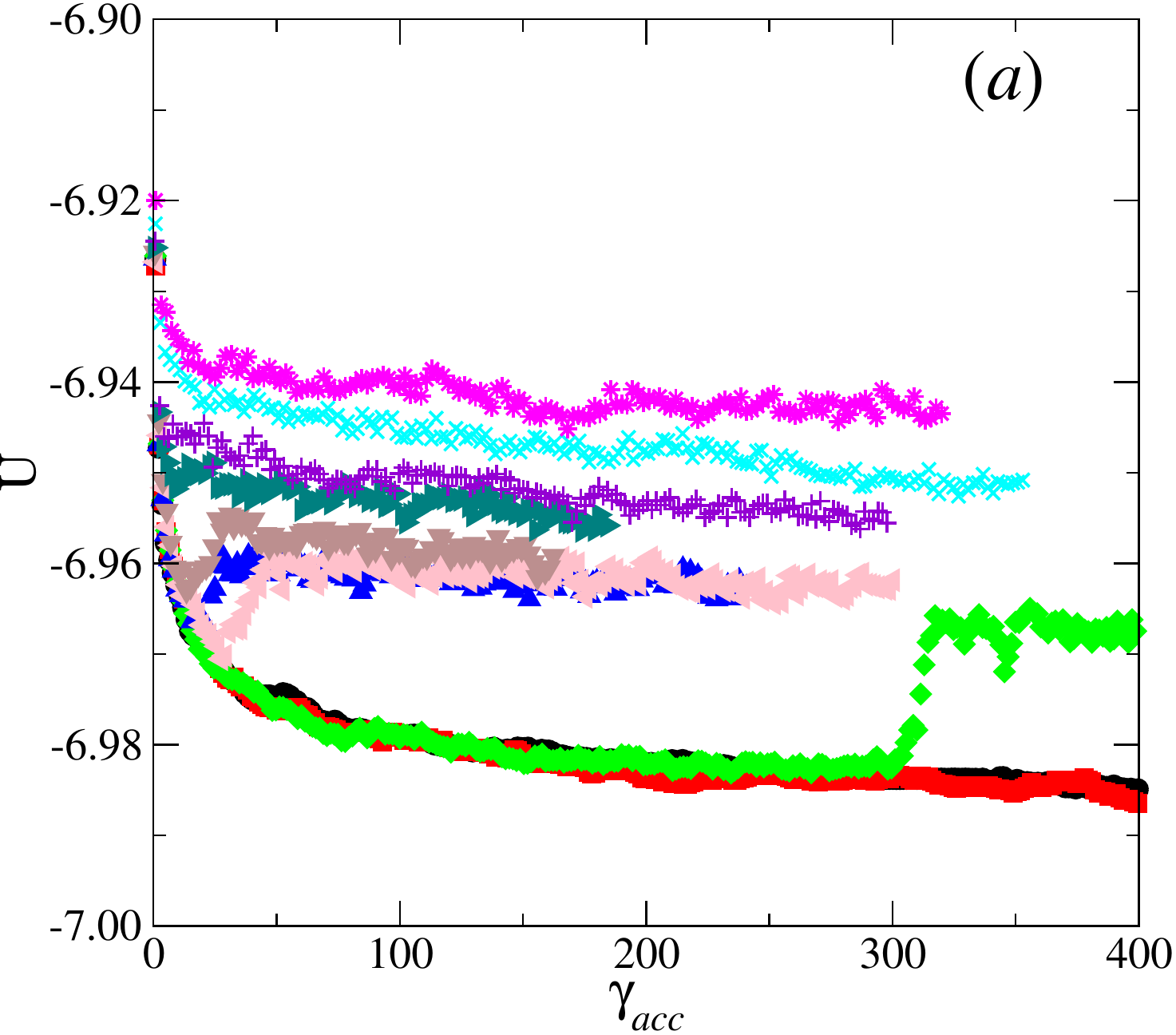} 
\includegraphics[scale=.45]{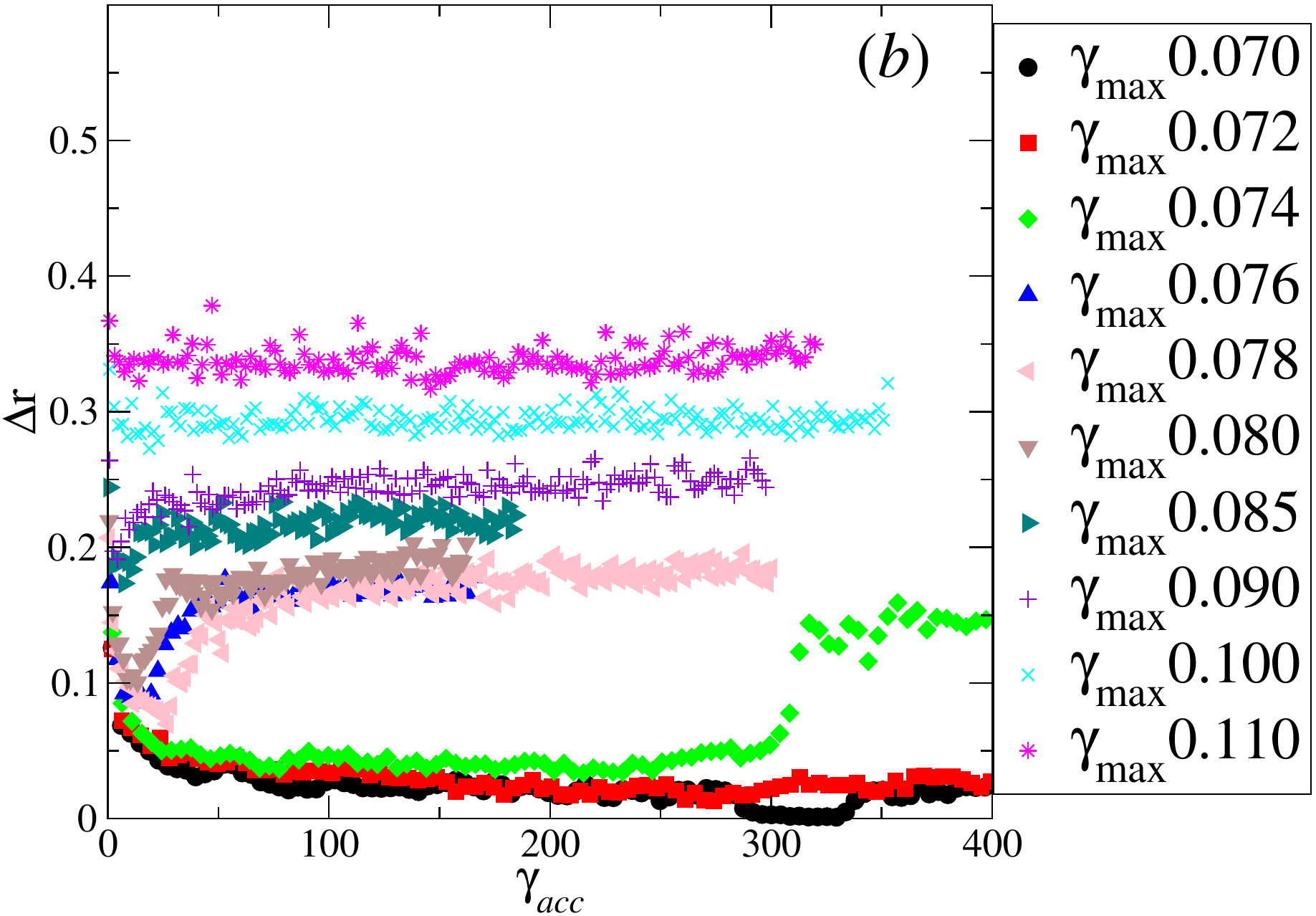} \\
\includegraphics[scale=.35]{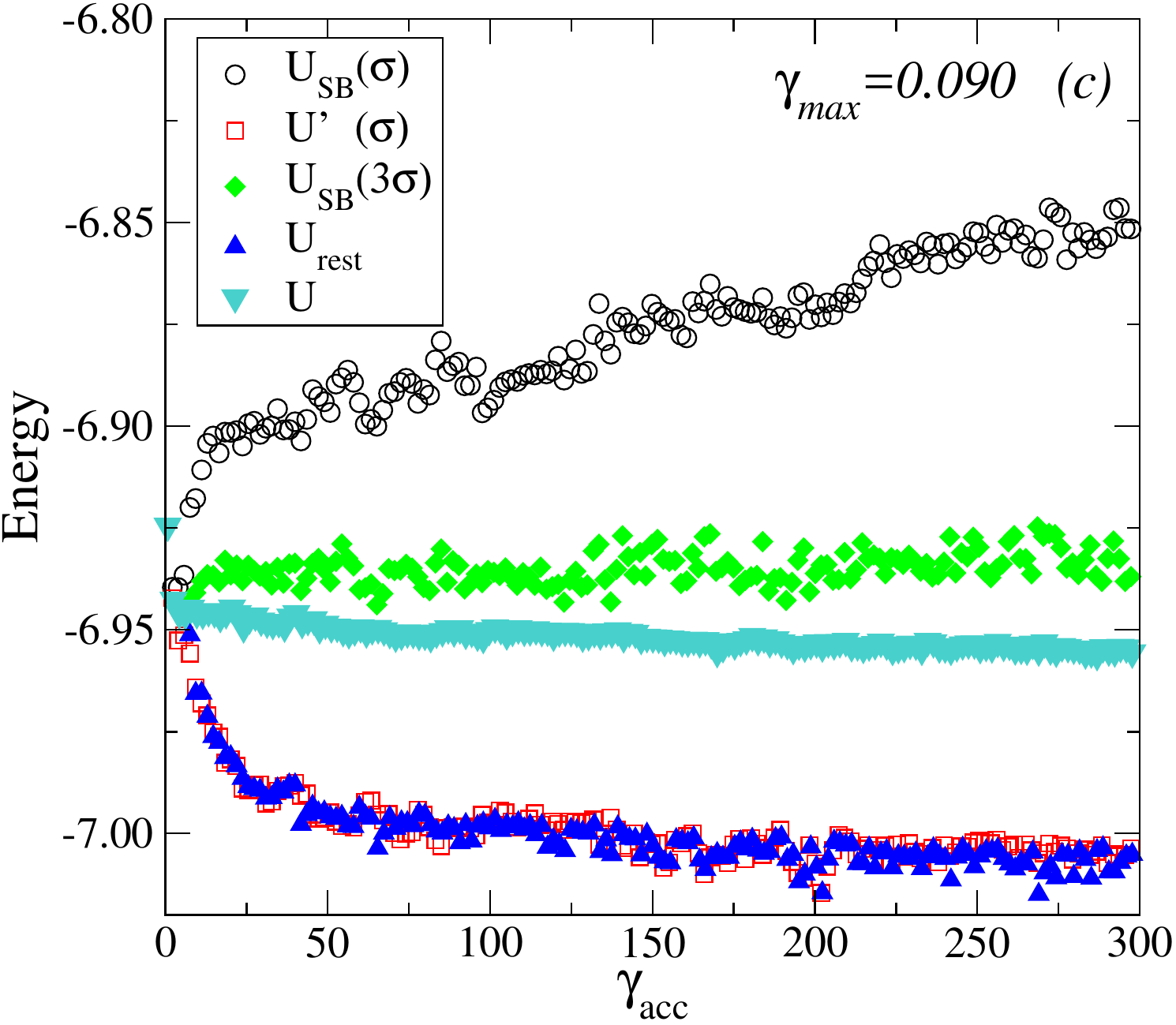} 
\includegraphics[scale=.35]{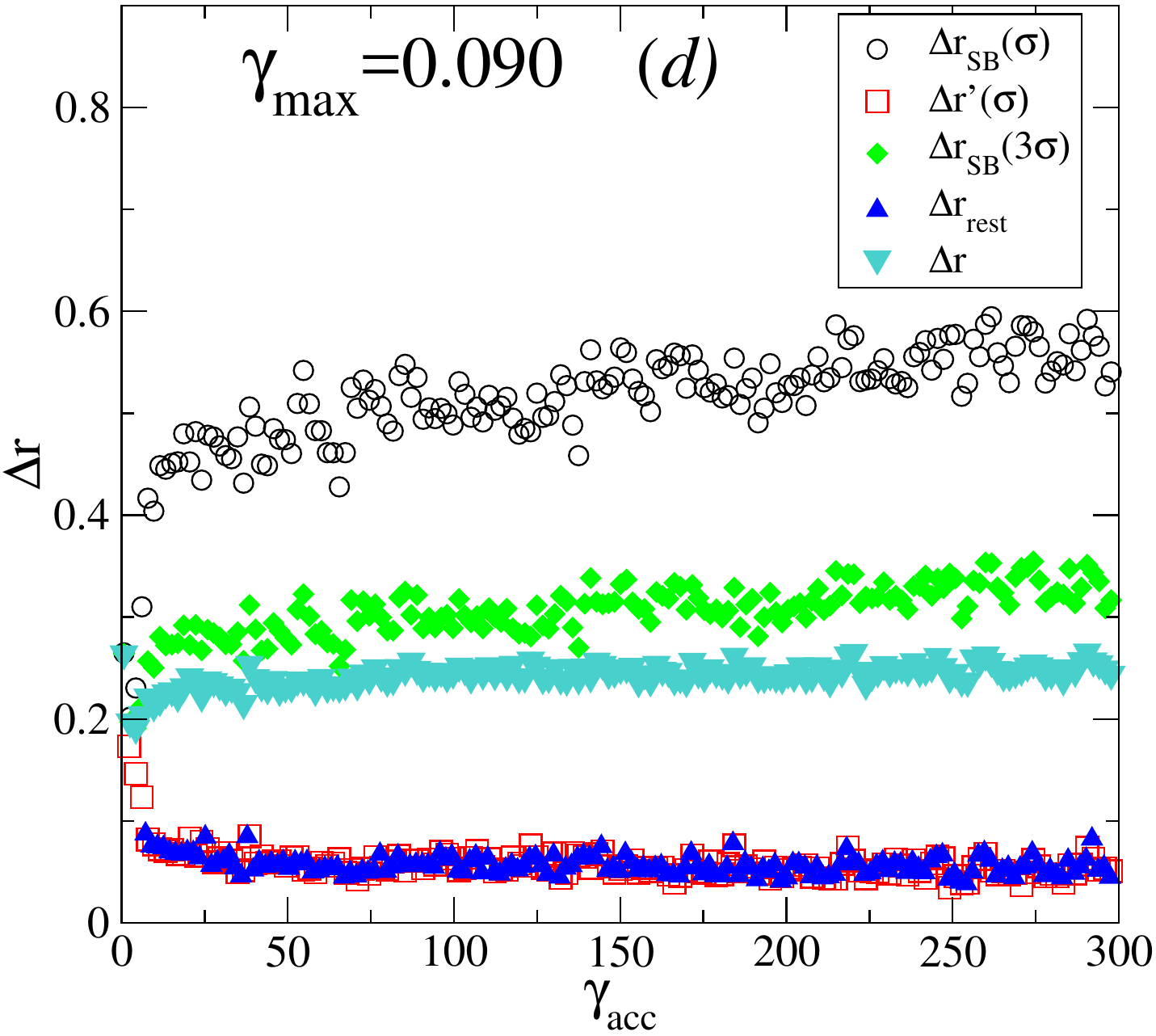} 
\includegraphics[scale=.35]{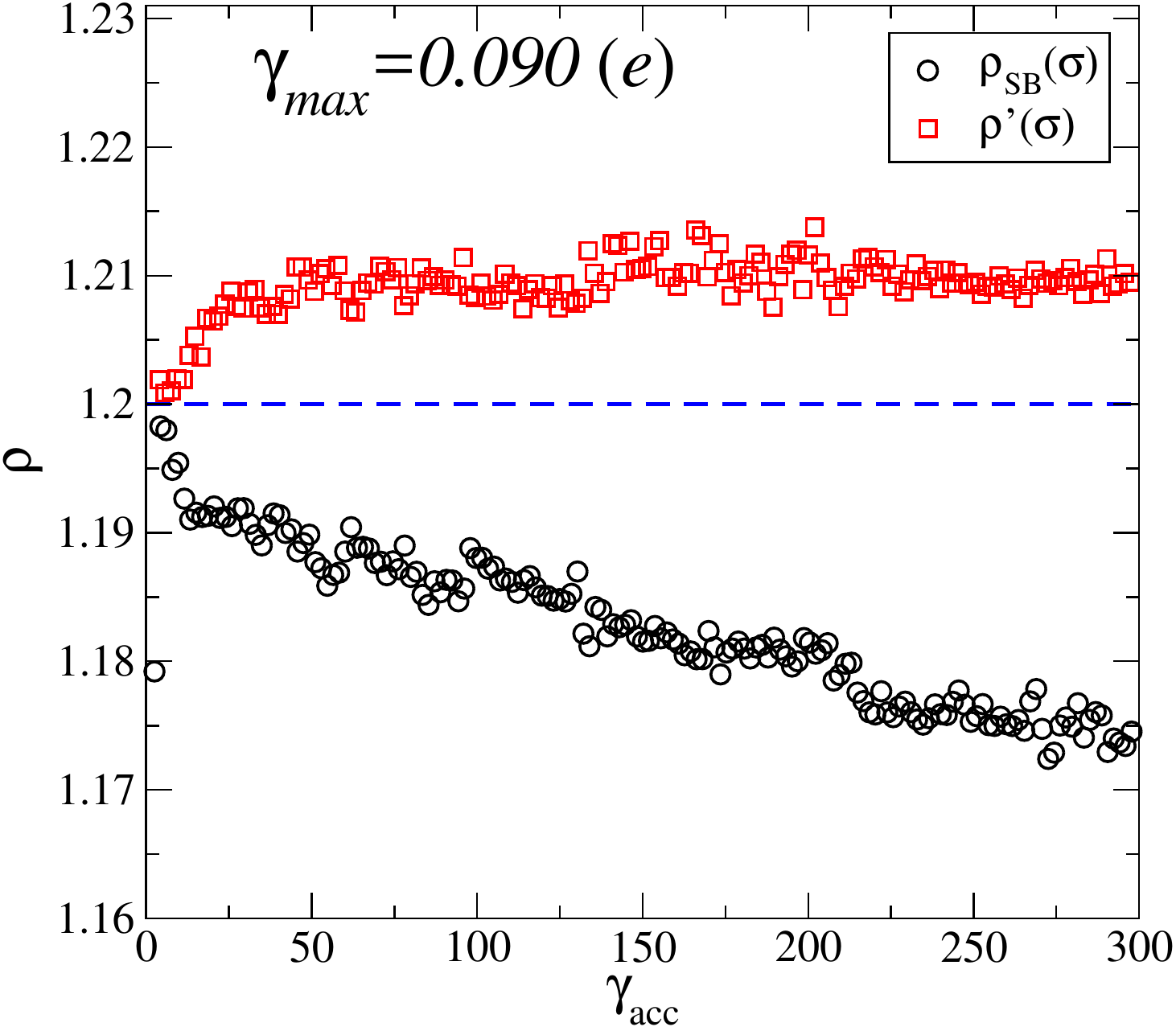} 
\caption{ {(\emph a)} The potential energy per particle ($U$) of stroboscopic configurations and {(\emph b)} averaged particle displacement ($\Delta r$) after one deformation cycle as a function of accumulated strain ($\gamma_{acc.}$) for various strain amplitude. For $\gamma_{max}< \gamma_y$, the system remains annealed and corresponding single particle displacements are small. The non-monotonic behaviour at higher amplitudes arises from shear banding. 
{(\emph c)} The potential energy and {(\emph d)} averaged particle displacement for ($i$) the centre of the shear band ($U_{SB}(\sigma)~\&~\Delta r_{SB}(\sigma)$), $ii$) slab farthest away ($U'(\sigma)~\&~\Delta r'(\sigma)$), ($iii$) most of the shear band ($U_{SB}(3\sigma)~\&~\Delta r_{SB}(3\sigma)$),   $iv$) therest of the system ($U_{rest}~\&~\Delta r_{rest}$) and ($v$) the entire system ($U~\&~\Delta r$) as a function of accumulated strain. The system outside the shear band remains annealed even for large strain amplitudes above the yielding amplitude. ({\emph e}) The local density of the shear band decreases compared with the mean density of the system.}
\label{fig2}
\end{figure*}

We next consider a detailed analysis of the energies and displacements as a function of accumulated strain $\gamma_{acc} \equiv N_{cycles} \times 4 \gamma_{max}$ for a range of strain amplitudes spanning the yield strain amplitude. In Fig. \ref{fig2}, the potential energy, $\Delta r$, and  the density $\rho$ of the full system, as well as various sub-volumes (within and outside the shear band) as shown against $\gamma_{acc}$.
The initial sample corresponds in each case is an inherent structures obtained from liquid configurations a high temperature  ($T=1$), roughly corresponding to the highest inherent structure energy at fixed density, often referred to as the {\it top of the landscape}.  For the smaller amplitudes, the energy $U$ and displacements $\Delta r$ decrease monotonically, indicating considerable annealing, as previously observed \cite{leishangthem2017yielding}. For larger amplitudes, the variation with  $\gamma_{acc}$ is non-monotonic, and the sharp upward changes 
(seen most markedly for $\gamma_{max} = 0.074$) indicate the onset of shear banding. Energy and  $\Delta r$ values shown as averages within and outside the shear band for $\gamma_{max} = 0.09$ in  Fig. \ref{fig2} (c), (d) reveal that the upward changes arise within the shear band  ($U_{SB}(3 \sigma)$ and $U_{SB}(\sigma)$, $\Delta r_{SB}(3\sigma)$ and $\Delta r_{SB}(\sigma)$) , whereas outside ( $U'_{SB}(\sigma)$, $U{rest}$,  $\Delta r^{'}(\sigma)$, $\Delta r_{rest}$), continued annealing is revealed by the monotonic decrease of the energy and displacements. Fig. \ref{fig2} (e) shows densities within and outside the shear band, displaying a densification outside the shear band, whereas within, the density shows significant  reduction.

\begin{figure*}[]
\centering
\includegraphics[scale=.50]{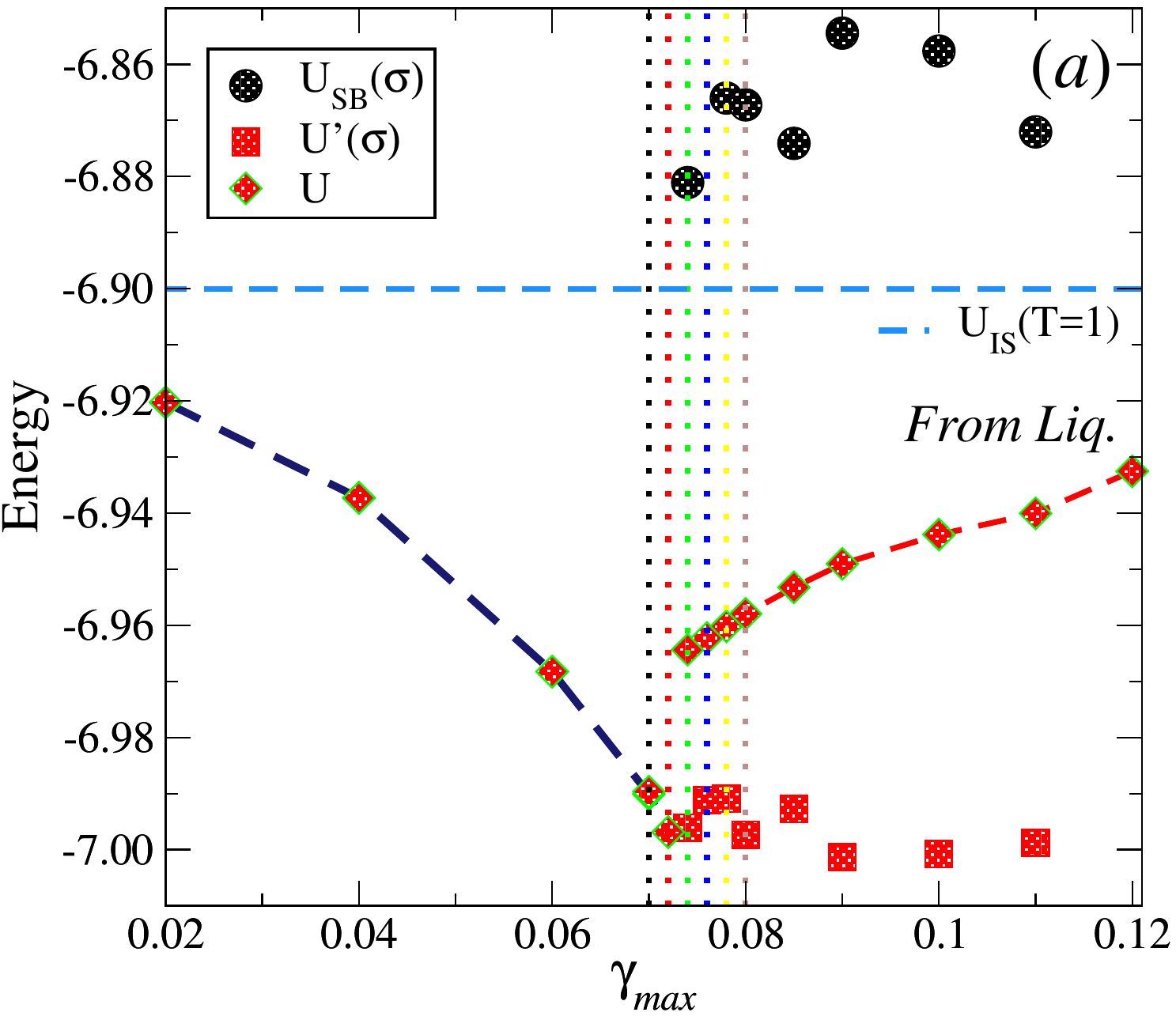} \hspace {-2mm}
\includegraphics[scale=.50]{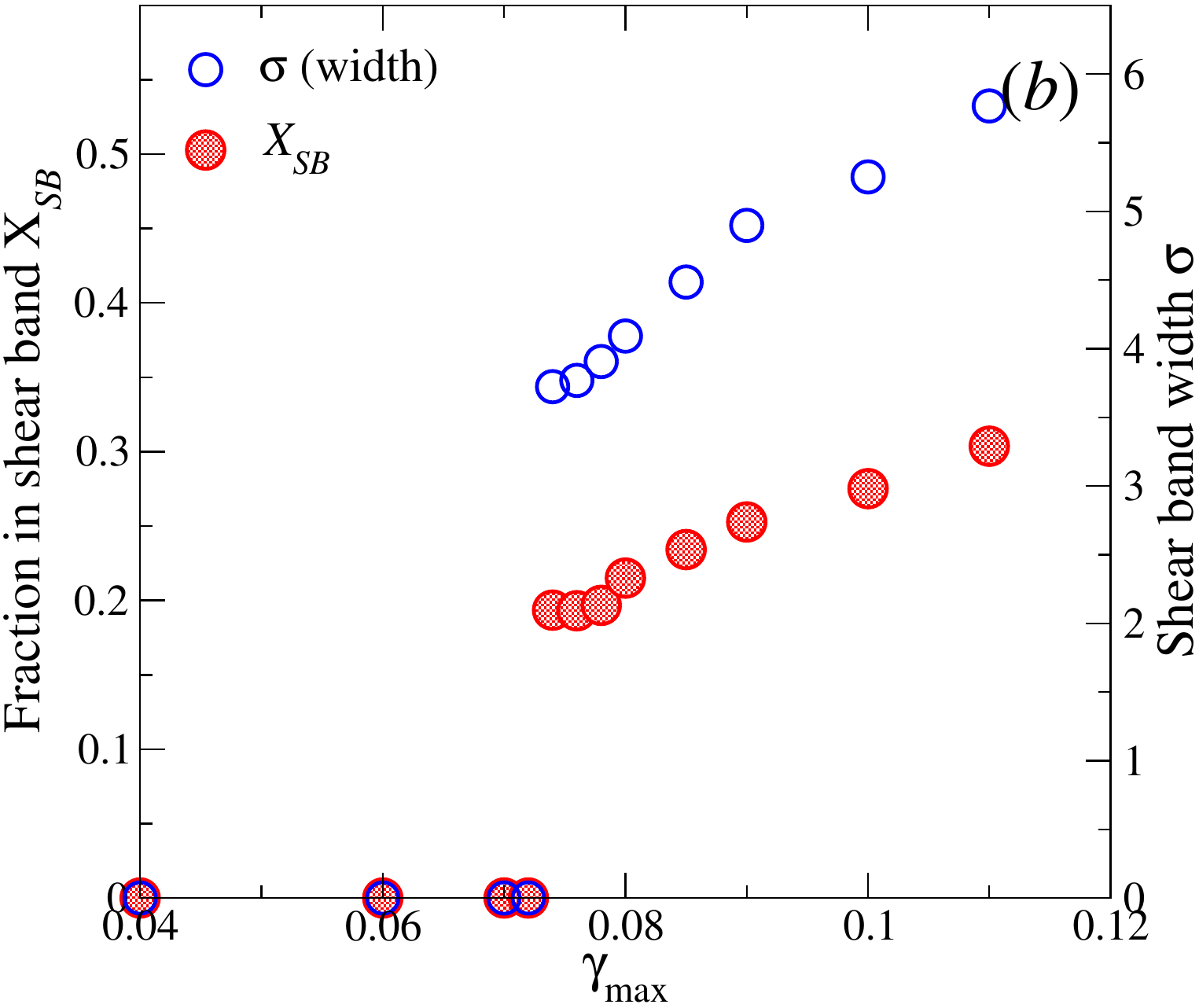} 
\caption{{(\emph a)} The mean energy of the system, shear band, and outside (farthest $\sigma$-wide slab) are plotted against strain amplitude. The mean energy changes discontinuously across the yielding transition. 
The system outside the shear band is increasingly better annealed for amplitudes even beyond the yield amplitude, whereas particles in the shear band have energies higher than the initial inherent structure energies, owing to a reduction in density.
The region between the strain amplitude $0.07$ to $0.08$ is highlighted by equally spaced lines, with the interval of strain $0.002$.   
{(\emph b)} The thickness of band ($\sigma$) and fraction of particles ($X_{SB}$, within $\sigma$) of the shear band change from zero discontinuously to finite values at the yielding point.}
\label{fig3}
\end{figure*}

We next show ($i$) the mean potential energy,
   ($ii$) the width of the shear band,  and ($iii$) the fraction of particles
   within the shear band, in  Fig. \ref{fig3}. In Fig. \ref{fig3}(a)  we show the mean potential energy
    of the total system as a function of strain amplitude, which changes discontinuously 
    across the yielding amplitude. Such behaviour has been described as rejuvenation  \cite{Lacks2004a,Fiocco2013,leishangthem2017yielding} but the energy values within and outside the shear band make it clear that while the part of the system within the shear band rejuvenates (attains higher energies), the rest of the system continues to anneal.  The jump in the mean potential energy is coincident with the coming into existence of a shear band with finite width (with core width $ 2 \sigma \approx 7.5 \sigma_{AA}$ at the yielding strain).  Fig. \ref{fig3} (b) shows the width of the shear band ($\sigma$) and the fraction of particles within the shear band, underscoring the discontinuous nature of the transition.

\begin{figure*}[htp]
\centering{}
\includegraphics[scale=.49]{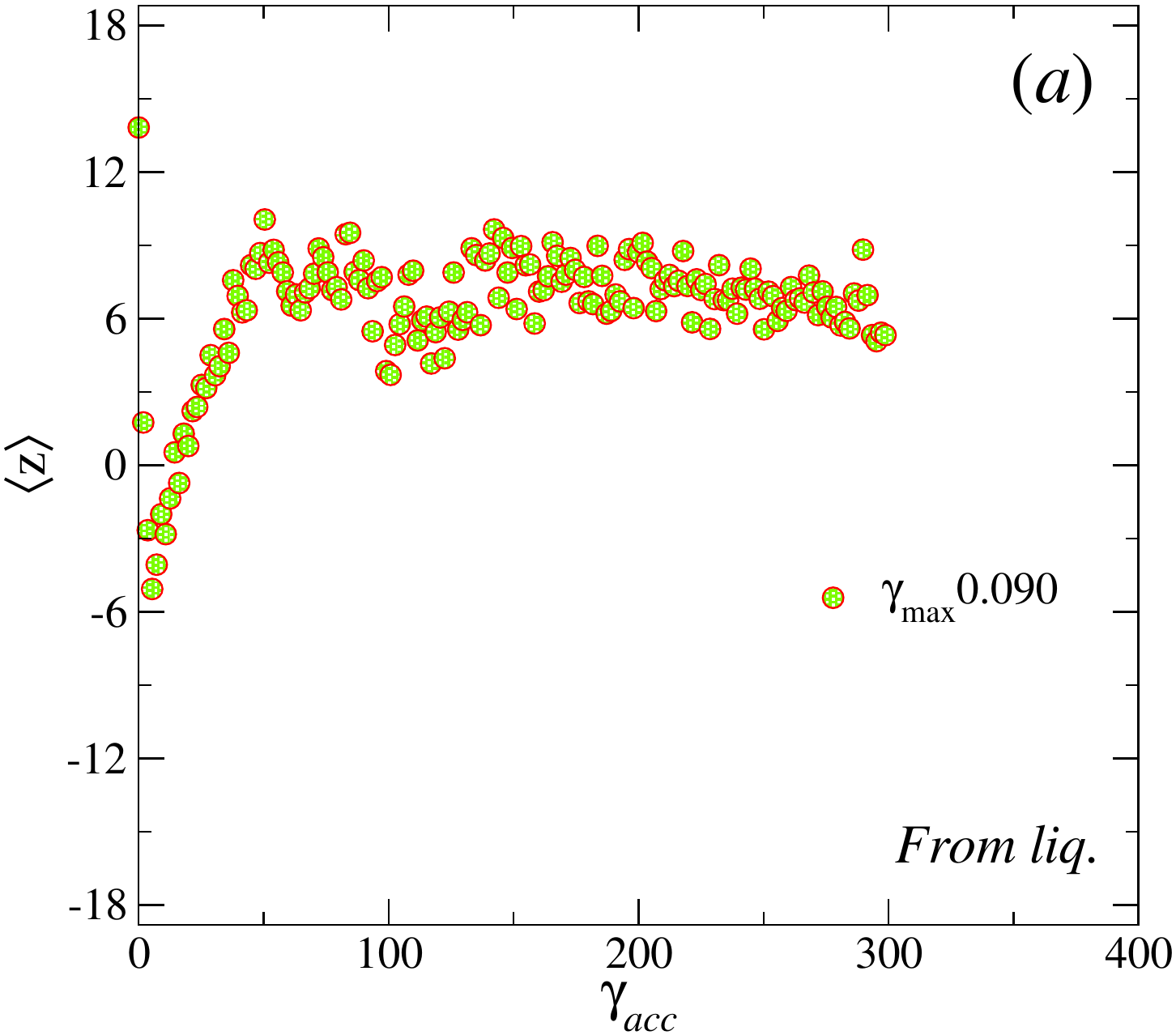}    
\includegraphics[scale=.49]{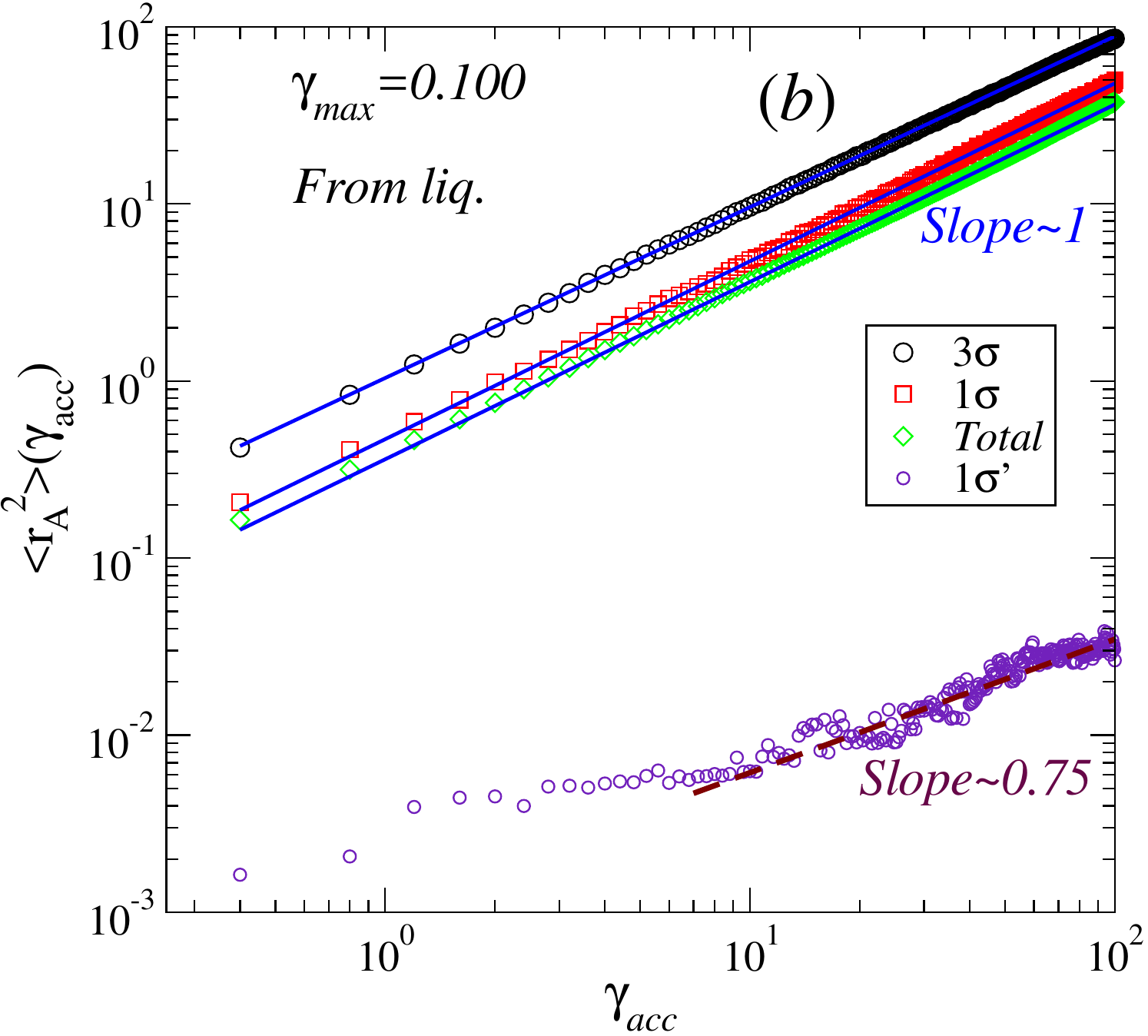}  
\includegraphics[scale=.49]{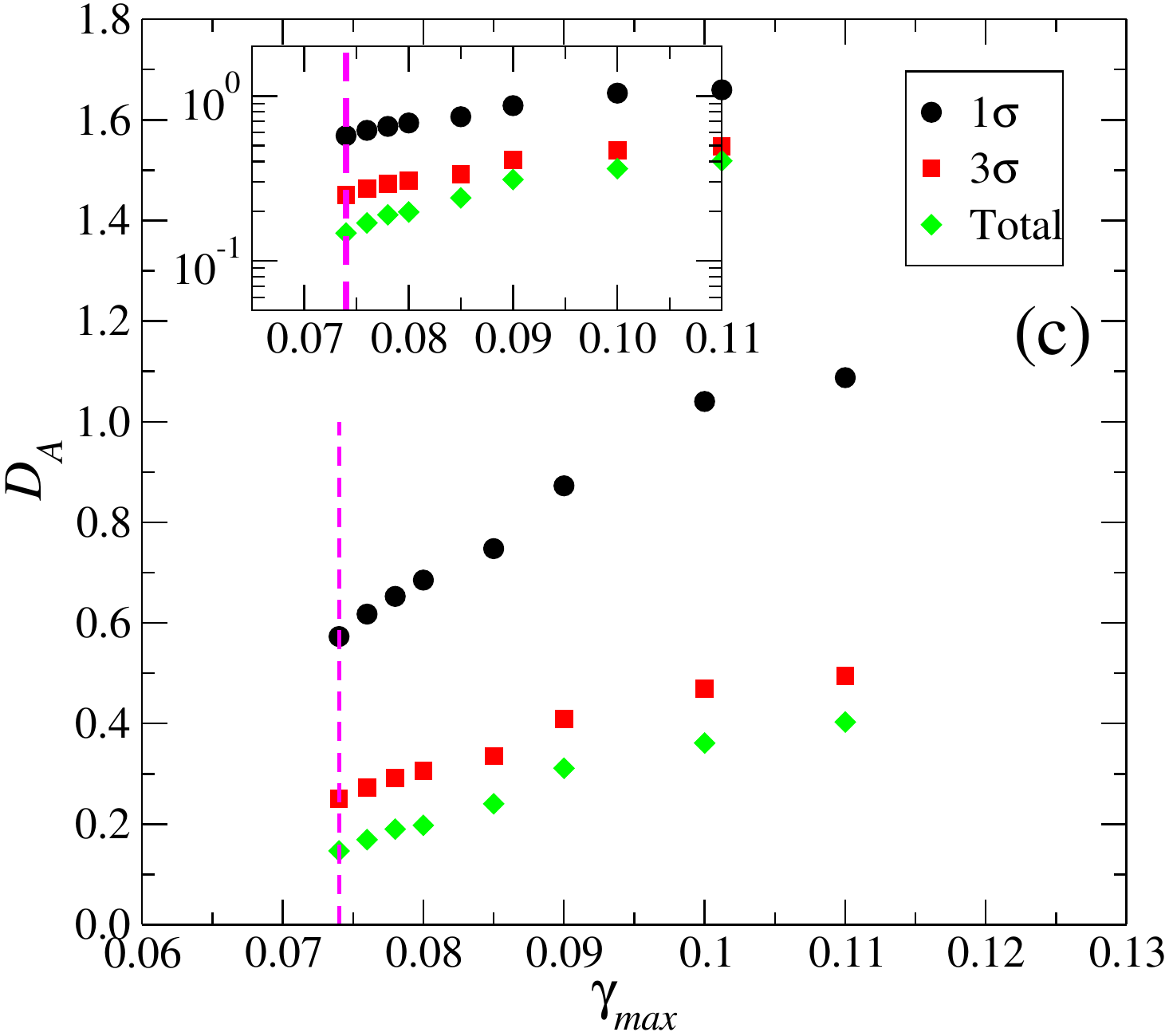}               
\includegraphics[scale=.49]{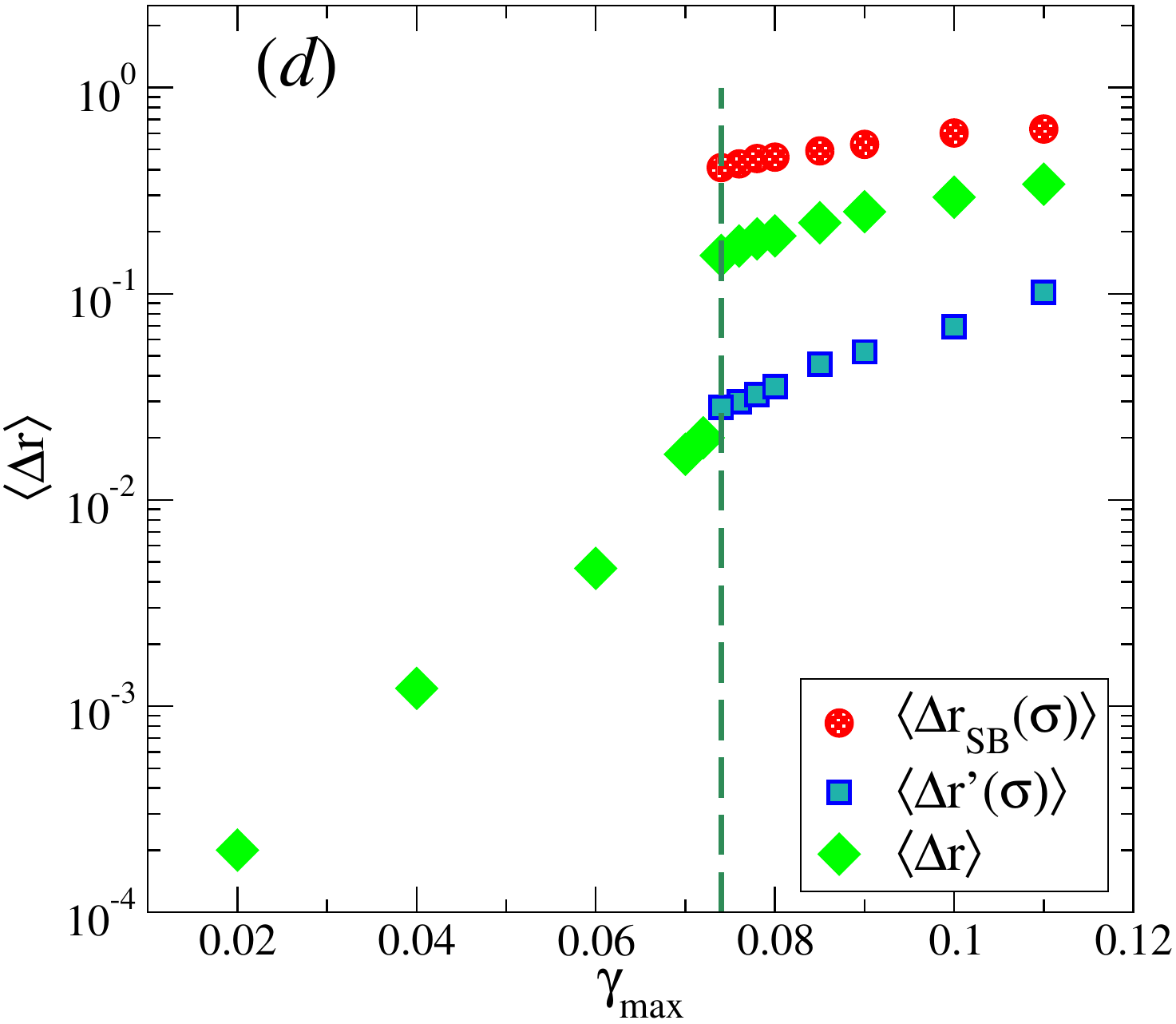}         
\caption{
{(\emph a)} The mean position of the shear band, $\langle z\rangle$, showing large scale variation with accumulated strain $\gamma_{acc}$. 
{(\emph b)}The mean squared displacement  of  ``$A$" type of particles for the full system, particles within the shear band ($\sigma$, $3 \sigma$) show linear variation with  $\gamma_{acc}$ whereas at the slab farthest away ($1 \sigma^{'}$) it is sub-diffusive. 
{(\emph c)} The diffusion coefficient of particles of type ``$A$" for the centre of the shear band ($\sigma$), most of the shear band ($3\sigma$) and of the total system, showing  discontinuous change across the yielding point. The inset shows the same data on a semi-log scale. 
{(\emph d)}   The steady state values of $\Delta r$ for the total system ($\langle \Delta r\rangle$), shear band ($\langle \Delta r_{SB}(\sigma) \rangle$), and the slab of thickness $\sigma$ farthest away from the shear band ($\langle \Delta r^{'} (\sigma)\rangle$), plotted against applied strain amplitudes. The $\langle \Delta r\rangle$ of the full system and the shear band change in a discontinuous manner across the yielding amplitude. $\Delta r$ values outside show continuous variation with values below the yielding amplitude.}
\label{fig4}
\end{figure*}

Fig. \ref{fig4}(a) shows the mean position $\langle z\rangle$ of the 
shear band for $\gamma_{max}=0.090$. The mean position shows movement over distances comparable to the dimensions of the simulated system, lending credence to the characterisation of the state of the system above yielding as {\it ergodic} \cite{Fiocco2013}, which may be doubted in the presence of shear banding. We calculate the mean squared displacements $\langle r^{2}_{A}\rangle$  (for ``$A$" type 
of species) in the steady state for the entire system, $\sigma$, $3\sigma$, 
and the slab of thickness $\sigma$ farthest from the shear band, for a range of strain amplitudes. 
As shown in Fig. \ref{fig4} (b), data for the  entire system and the shear band 
show  diffusive behaviour, whereas for the slab of thickness $\sigma$ farthest from the shear band  (annealed region) the behaviour is sub-diffusive.
The diffusion coefficients ($D_{A}$)  estimated 
from the fit function, $\langle r^{2}_{A}\rangle(\gamma_{acc})=D_{A}\gamma_{acc}$, and shown in  Fig. \ref{fig4} (c), indicate that the diffusion coefficients change discontinuously from zero to a finite value across the yielding transition, consistently with the findings in \cite{kawasaki2016macroscopic}.
Such discontinuous behaviour is also seen in the behaviour of $\langle \Delta r \rangle$, for which  (Fig. \ref{fig4} (d)) a decomposition into values within and outside the shear band shows that the discontinuity arises from the emergence of the shear band.

\begin{figure*}[htp]
\centering 
\includegraphics[scale=.40]{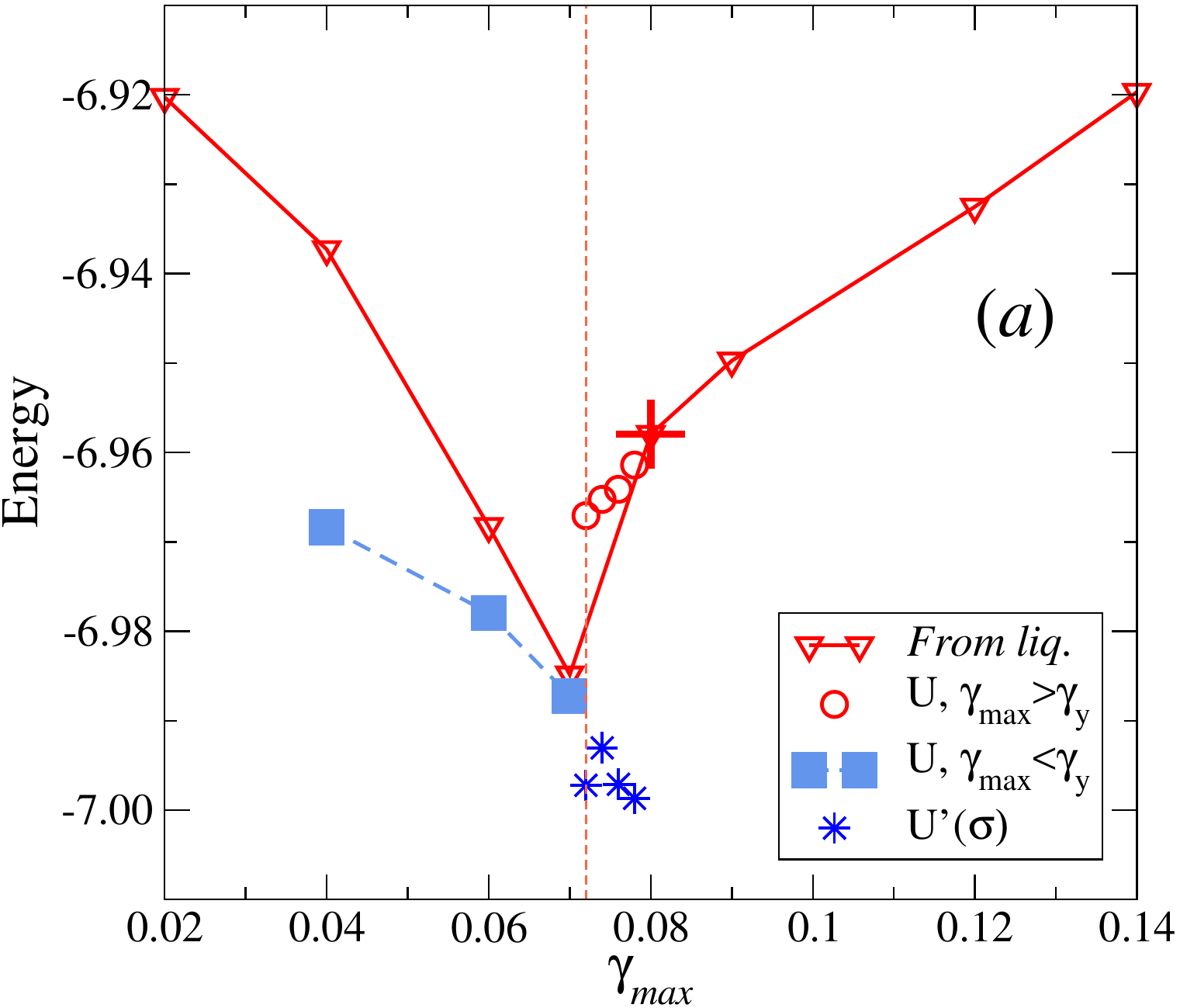}
\includegraphics[scale=.40]{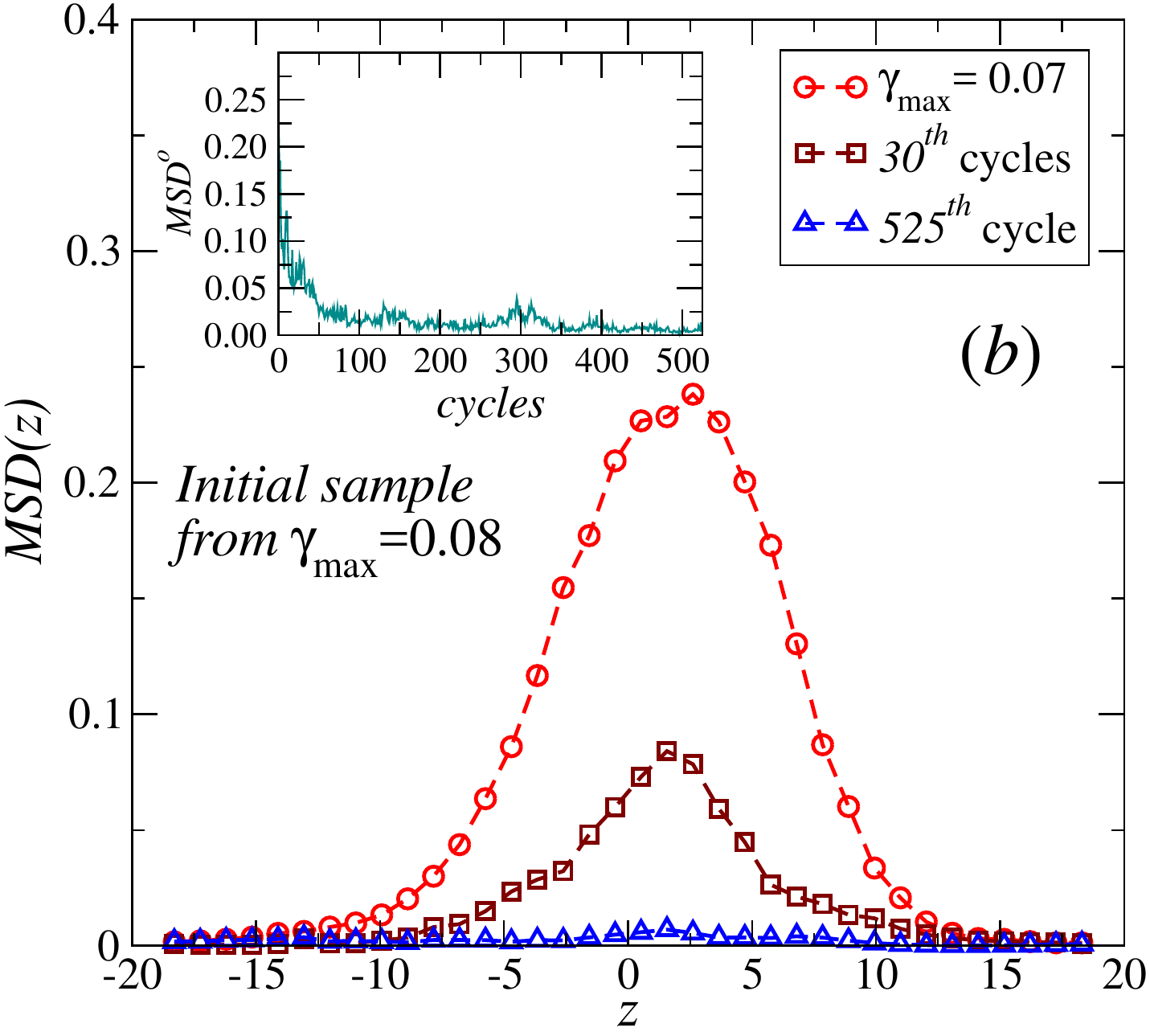}
\caption{{(\emph a)} Average energies obtained when an initial (shear banded) sample from the the steady state at $\gamma_{max}=0.08$, is deformed for a range of amplitudes above and below the yielding amplitude.  Below the yield amplitude ($<\gamma_y$), shear band anneals out (as seen in (b)), leading to energies below those obtained with inherent structures from the high temperature liquid at $T = 1$ are used as initial configurations (red triangles shown as reference). Above yielding, energies within and outside the shear band are different.  ({\emph b}) $MSD(z)$ along the $z$ direction at different cycles showing the annealing out of the shear band. Inset shows the amplitude $MSD^{o}$ for the shear band {\it vs.} accumulated strain. }
\label{fig5}
\end{figure*} 

We finally show, in Fig. \ref{fig5}, results when an initially shear banded sample (prepared at  $\gamma_{max}=0.08$) is subjected to cyclic deformation for a range of amplitudes above and below the yielding amplitude. When  $\gamma_{max}$ is below the yielding amplitude, the shear band is gradually annealed out, and the system achieves energies that are lower than those obtained when  inherent structures from the high temperature liquid at $T = 1$ are used as initial configurations (see Fig. \ref{fig5} (a)). This is clearly demonstrated in  Fig. \ref{fig5} (b) through $MSD(z)$ shown at different numbers of cycles, which reveal that the amplitude of the $MSD$ gradually diminishes and becomes negligible. In other words, by performing cyclic deformation at an amplitude below yielding, even an initially shear banded state reaches a final state that is homogeneous and well annealed.

In summary, we have investigated strain localisation and formation of shear bands accompanying yielding in glasses subjected to athermal cyclic shear deformation. Shear bands with a thickness of several atomic diameters emerge discontinuously when the yielding amplitude of deformation is exceeded, consistently with, and illustrating, the descriptions of yielding as a discontinuous transition for well annealed glasses \cite{shi2005strain,Itamar2016yielding,Urbani2017b,parisi2017shear,leishangthem2017yielding,Ozawa2018a,Popovic2018a}.  We show that shear banded structures simultaneously show features of rejuvenation (inside the shear band) and annealing or aging (outside the shear band), and attain homogeneous states when subjected to cyclic deformation at amplitudes below the yielding amplitude, even when the initial state is shear banded. Some aspects of this behaviour have been observed in a soft glassy rheology model \cite{Radhakrishnan2016b} of oscillatory shear deformation, and are expected to be of importance also in comprehending the role of annealing in transient shear banding under uniform shear ({\it e. g.}, \cite{vasisht2017emergence}), and other situations, thus meriting efforts to rationalise them also through other approaches. 


\bibliography{mybib}{}

\begin{thebibliography}{46}%
\makeatletter
\providecommand \@ifxundefined [1]{%
 \@ifx{#1\undefined}
}%
\providecommand \@ifnum [1]{%
 \ifnum #1\expandafter \@firstoftwo
 \else \expandafter \@secondoftwo
 \fi
}%
\providecommand \@ifx [1]{%
 \ifx #1\expandafter \@firstoftwo
 \else \expandafter \@secondoftwo
 \fi
}%
\providecommand \natexlab [1]{#1}%
\providecommand \enquote  [1]{``#1''}%
\providecommand \bibnamefont  [1]{#1}%
\providecommand \bibfnamefont [1]{#1}%
\providecommand \citenamefont [1]{#1}%
\providecommand \href@noop [0]{\@secondoftwo}%
\providecommand \href [0]{\begingroup \@sanitize@url \@href}%
\providecommand \@href[1]{\@@startlink{#1}\@@href}%
\providecommand \@@href[1]{\endgroup#1\@@endlink}%
\providecommand \@sanitize@url [0]{\catcode `\\12\catcode `\$12\catcode
  `\&12\catcode `\#12\catcode `\^12\catcode `\_12\catcode `\%12\relax}%
\providecommand \@@startlink[1]{}%
\providecommand \@@endlink[0]{}%
\providecommand \url  [0]{\begingroup\@sanitize@url \@url }%
\providecommand \@url [1]{\endgroup\@href {#1}{\urlprefix }}%
\providecommand \urlprefix  [0]{URL }%
\providecommand \Eprint [0]{\href }%
\providecommand \doibase [0]{http://dx.doi.org/}%
\providecommand \selectlanguage [0]{\@gobble}%
\providecommand \bibinfo  [0]{\@secondoftwo}%
\providecommand \bibfield  [0]{\@secondoftwo}%
\providecommand \translation [1]{[#1]}%
\providecommand \BibitemOpen [0]{}%
\providecommand \bibitemStop [0]{}%
\providecommand \bibitemNoStop [0]{.\EOS\space}%
\providecommand \EOS [0]{\spacefactor3000\relax}%
\providecommand \BibitemShut  [1]{\csname bibitem#1\endcsname}%
\let\auto@bib@innerbib\@empty
\bibitem [{\citenamefont {Schuh}\ \emph {et~al.}(2007)\citenamefont {Schuh},
  \citenamefont {Hufnagel},\ and\ \citenamefont {Ramamurty}}]{Schuh2007a}%
  \BibitemOpen
  \bibfield  {author} {\bibinfo {author} {\bibfnamefont {C.~A.}\ \bibnamefont
  {Schuh}}, \bibinfo {author} {\bibfnamefont {T.~C.}\ \bibnamefont {Hufnagel}},
  \ and\ \bibinfo {author} {\bibfnamefont {U.}~\bibnamefont {Ramamurty}},\
  }\href {\doibase 10.1016/j.actamat.2007.01.052} {\bibfield  {journal}
  {\bibinfo  {journal} {Acta Mater.}\ }\textbf {\bibinfo {volume} {55}},\
  \bibinfo {pages} {4067} (\bibinfo {year} {2007})}\BibitemShut {NoStop}%
\bibitem [{\citenamefont {Schall}\ and\ \citenamefont {van
  Hecke}(2010)}]{Schall2010}%
  \BibitemOpen
  \bibfield  {author} {\bibinfo {author} {\bibfnamefont {P.}~\bibnamefont
  {Schall}}\ and\ \bibinfo {author} {\bibfnamefont {M.}~\bibnamefont {van
  Hecke}},\ }\href {\doibase 10.1146/annurev-fluid-121108-145544} {\bibfield
  {journal} {\bibinfo  {journal} {Annual Review of Fluid Mechanics}\ }\textbf
  {\bibinfo {volume} {42}},\ \bibinfo {pages} {67} (\bibinfo {year}
  {2010})}\BibitemShut {NoStop}%
\bibitem [{\citenamefont {Falk}\ and\ \citenamefont {Langer}(2011)}]{Falk2010}%
  \BibitemOpen
  \bibfield  {author} {\bibinfo {author} {\bibfnamefont {M.~L.}\ \bibnamefont
  {Falk}}\ and\ \bibinfo {author} {\bibfnamefont {J.}~\bibnamefont {Langer}},\
  }\href {\doibase 10.1146/annurev-conmatphys-062910-140452} {\bibfield
  {journal} {\bibinfo  {journal} {Annual Review of Condensed Matter Physics}\
  }\textbf {\bibinfo {volume} {2}},\ \bibinfo {pages} {353} (\bibinfo {year}
  {2011})}\BibitemShut {NoStop}%
\bibitem [{\citenamefont {Bonn}\ \emph {et~al.}(2017)\citenamefont {Bonn},
  \citenamefont {Denn}, \citenamefont {Berthier}, \citenamefont {Divoux},\ and\
  \citenamefont {Manneville}}]{Bonn2017c}%
  \BibitemOpen
  \bibfield  {author} {\bibinfo {author} {\bibfnamefont {D.}~\bibnamefont
  {Bonn}}, \bibinfo {author} {\bibfnamefont {M.~M.}\ \bibnamefont {Denn}},
  \bibinfo {author} {\bibfnamefont {L.}~\bibnamefont {Berthier}}, \bibinfo
  {author} {\bibfnamefont {T.}~\bibnamefont {Divoux}}, \ and\ \bibinfo {author}
  {\bibfnamefont {S.}~\bibnamefont {Manneville}},\ }\href@noop {} {\bibfield
  {journal} {\bibinfo  {journal} {Rev. Mod. Phys.}\ }\textbf {\bibinfo {volume}
  {89}} (\bibinfo {year} {2017})}\BibitemShut {NoStop}%
\bibitem [{\citenamefont {Nicolas}\ \emph {et~al.}(2017)\citenamefont
  {Nicolas}, \citenamefont {Ferrero}, \citenamefont {Martens},\ and\
  \citenamefont {Barrat}}]{Nicolas2017a}%
  \BibitemOpen
  \bibfield  {author} {\bibinfo {author} {\bibfnamefont {A.}~\bibnamefont
  {Nicolas}}, \bibinfo {author} {\bibfnamefont {E.~E.}\ \bibnamefont
  {Ferrero}}, \bibinfo {author} {\bibfnamefont {K.}~\bibnamefont {Martens}}, \
  and\ \bibinfo {author} {\bibfnamefont {J.-L.}\ \bibnamefont {Barrat}},\
  }\href {http://arxiv.org/abs/1708.09194} {\  (\bibinfo {year} {2017})},\
  \Eprint {http://arxiv.org/abs/1708.09194} {arXiv:1708.09194} \BibitemShut
  {NoStop}%
\bibitem [{\citenamefont {Chester}\ and\ \citenamefont
  {Chester}(1998)}]{chester1998ultracataclasite}%
  \BibitemOpen
  \bibfield  {author} {\bibinfo {author} {\bibfnamefont {F.~M.}\ \bibnamefont
  {Chester}}\ and\ \bibinfo {author} {\bibfnamefont {J.~S.}\ \bibnamefont
  {Chester}},\ }\href {\doibase https://doi.org/10.1016/S0040-1951(98)00121-8}
  {\bibfield  {journal} {\bibinfo  {journal} {Tectonophysics}\ }\textbf
  {\bibinfo {volume} {295}},\ \bibinfo {pages} {199} (\bibinfo {year}
  {1998})}\BibitemShut {NoStop}%
\bibitem [{\citenamefont {Pekarskaya}\ \emph {et~al.}(2001)\citenamefont
  {Pekarskaya}, \citenamefont {Kim},\ and\ \citenamefont
  {Johnson}}]{pekarskaya2001situ}%
  \BibitemOpen
  \bibfield  {author} {\bibinfo {author} {\bibfnamefont {E.}~\bibnamefont
  {Pekarskaya}}, \bibinfo {author} {\bibfnamefont {C.}~\bibnamefont {Kim}}, \
  and\ \bibinfo {author} {\bibfnamefont {W.}~\bibnamefont {Johnson}},\ }\href
  {\doibase 10.1557/JMR.2001.0344} {\bibfield  {journal} {\bibinfo  {journal}
  {Journal of Materials Research}\ }\textbf {\bibinfo {volume} {16}},\ \bibinfo
  {pages} {2513} (\bibinfo {year} {2001})}\BibitemShut {NoStop}%
\bibitem [{\citenamefont {Cohen}\ \emph {et~al.}(2006)\citenamefont {Cohen},
  \citenamefont {Davidovitch}, \citenamefont {Schofield}, \citenamefont
  {Brenner},\ and\ \citenamefont {Weitz}}]{cohen2006slip}%
  \BibitemOpen
  \bibfield  {author} {\bibinfo {author} {\bibfnamefont {I.}~\bibnamefont
  {Cohen}}, \bibinfo {author} {\bibfnamefont {B.}~\bibnamefont {Davidovitch}},
  \bibinfo {author} {\bibfnamefont {A.~B.}\ \bibnamefont {Schofield}}, \bibinfo
  {author} {\bibfnamefont {M.~P.}\ \bibnamefont {Brenner}}, \ and\ \bibinfo
  {author} {\bibfnamefont {D.~A.}\ \bibnamefont {Weitz}},\ }\href {\doibase
  10.1103/PhysRevLett.97.215502} {\bibfield  {journal} {\bibinfo  {journal}
  {Phys. Rev. Lett.}\ }\textbf {\bibinfo {volume} {97}},\ \bibinfo {pages}
  {215502} (\bibinfo {year} {2006})}\BibitemShut {NoStop}%
\bibitem [{\citenamefont {Uhl}\ \emph {et~al.}(2015)\citenamefont {Uhl},
  \citenamefont {Pathak}, \citenamefont {Schorlemmer}, \citenamefont {Liu},
  \citenamefont {Swindeman}, \citenamefont {Brinkman}, \citenamefont {LeBlanc},
  \citenamefont {Tsekenis}, \citenamefont {Friedman}, \citenamefont
  {Behringer}, \citenamefont {Denisov}, \citenamefont {Schall}, \citenamefont
  {Gu}, \citenamefont {Wright}, \citenamefont {Hufnagel}, \citenamefont
  {Jennings}, \citenamefont {Greer}, \citenamefont {Liaw}, \citenamefont
  {Becker}, \citenamefont {Dresen},\ and\ \citenamefont {Dahmen}}]{Uhl2015b}%
  \BibitemOpen
  \bibfield  {author} {\bibinfo {author} {\bibfnamefont {J.~T.}\ \bibnamefont
  {Uhl}}, \bibinfo {author} {\bibfnamefont {S.}~\bibnamefont {Pathak}},
  \bibinfo {author} {\bibfnamefont {D.}~\bibnamefont {Schorlemmer}}, \bibinfo
  {author} {\bibfnamefont {X.}~\bibnamefont {Liu}}, \bibinfo {author}
  {\bibfnamefont {R.}~\bibnamefont {Swindeman}}, \bibinfo {author}
  {\bibfnamefont {B.~A.}\ \bibnamefont {Brinkman}}, \bibinfo {author}
  {\bibfnamefont {M.}~\bibnamefont {LeBlanc}}, \bibinfo {author} {\bibfnamefont
  {G.}~\bibnamefont {Tsekenis}}, \bibinfo {author} {\bibfnamefont
  {N.}~\bibnamefont {Friedman}}, \bibinfo {author} {\bibfnamefont
  {R.}~\bibnamefont {Behringer}}, \bibinfo {author} {\bibfnamefont
  {D.}~\bibnamefont {Denisov}}, \bibinfo {author} {\bibfnamefont
  {P.}~\bibnamefont {Schall}}, \bibinfo {author} {\bibfnamefont
  {X.}~\bibnamefont {Gu}}, \bibinfo {author} {\bibfnamefont {W.~J.}\
  \bibnamefont {Wright}}, \bibinfo {author} {\bibfnamefont {T.}~\bibnamefont
  {Hufnagel}}, \bibinfo {author} {\bibfnamefont {A.}~\bibnamefont {Jennings}},
  \bibinfo {author} {\bibfnamefont {J.~R.}\ \bibnamefont {Greer}}, \bibinfo
  {author} {\bibfnamefont {P.~K.}\ \bibnamefont {Liaw}}, \bibinfo {author}
  {\bibfnamefont {T.}~\bibnamefont {Becker}}, \bibinfo {author} {\bibfnamefont
  {G.}~\bibnamefont {Dresen}}, \ and\ \bibinfo {author} {\bibfnamefont {K.~A.}\
  \bibnamefont {Dahmen}},\ }\href {\doibase 10.1038/srep16493} {\bibfield
  {journal} {\bibinfo  {journal} {Sci. Rep.}\ }\textbf {\bibinfo {volume}
  {5}},\ \bibinfo {pages} {1} (\bibinfo {year} {2015})}\BibitemShut {NoStop}%
\bibitem [{\citenamefont {H\'ebraud}\ and\ \citenamefont
  {Lequeux}(1998)}]{Hebraud1998}%
  \BibitemOpen
  \bibfield  {author} {\bibinfo {author} {\bibfnamefont {P.}~\bibnamefont
  {H\'ebraud}}\ and\ \bibinfo {author} {\bibfnamefont {F.}~\bibnamefont
  {Lequeux}},\ }\href {\doibase 10.1103/PhysRevLett.81.2934} {\bibfield
  {journal} {\bibinfo  {journal} {Phys. Rev. Lett.}\ }\textbf {\bibinfo
  {volume} {81}},\ \bibinfo {pages} {2934} (\bibinfo {year}
  {1998})}\BibitemShut {NoStop}%
\bibitem [{\citenamefont {Shi}\ and\ \citenamefont
  {Falk}(2005)}]{shi2005strain}%
  \BibitemOpen
  \bibfield  {author} {\bibinfo {author} {\bibfnamefont {Y.}~\bibnamefont
  {Shi}}\ and\ \bibinfo {author} {\bibfnamefont {M.~L.}\ \bibnamefont {Falk}},\
  }\href {\doibase 10.1103/PhysRevLett.95.095502} {\bibfield  {journal}
  {\bibinfo  {journal} {Phys. Rev. Lett.}\ }\textbf {\bibinfo {volume} {95}},\
  \bibinfo {pages} {095502} (\bibinfo {year} {2005})}\BibitemShut {NoStop}%
\bibitem [{\citenamefont {Maloney}\ and\ \citenamefont
  {Lema\^{\i}tre}(2006)}]{Maloney2006}%
  \BibitemOpen
  \bibfield  {author} {\bibinfo {author} {\bibfnamefont {C.~E.}\ \bibnamefont
  {Maloney}}\ and\ \bibinfo {author} {\bibfnamefont {A.}~\bibnamefont
  {Lema\^{\i}tre}},\ }\href {\doibase 10.1103/PhysRevE.74.016118} {\bibfield
  {journal} {\bibinfo  {journal} {Phys. Rev. E}\ }\textbf {\bibinfo {volume}
  {74}},\ \bibinfo {pages} {016118} (\bibinfo {year} {2006})}\BibitemShut
  {NoStop}%
\bibitem [{\citenamefont {Dahmen}\ \emph {et~al.}(2009)\citenamefont {Dahmen},
  \citenamefont {Ben-Zion},\ and\ \citenamefont {Uhl}}]{Dahmen2009}%
  \BibitemOpen
  \bibfield  {author} {\bibinfo {author} {\bibfnamefont {K.~A.}\ \bibnamefont
  {Dahmen}}, \bibinfo {author} {\bibfnamefont {Y.}~\bibnamefont {Ben-Zion}}, \
  and\ \bibinfo {author} {\bibfnamefont {J.~T.}\ \bibnamefont {Uhl}},\ }\href
  {\doibase 10.1103/PhysRevLett.102.175501} {\bibfield  {journal} {\bibinfo
  {journal} {Phys. Rev. Lett.}\ }\textbf {\bibinfo {volume} {102}},\ \bibinfo
  {pages} {175501} (\bibinfo {year} {2009})}\BibitemShut {NoStop}%
\bibitem [{\citenamefont {Manning}\ \emph {et~al.}(2009)\citenamefont
  {Manning}, \citenamefont {Daub}, \citenamefont {Langer},\ and\ \citenamefont
  {Carlson}}]{manning2009rate}%
  \BibitemOpen
  \bibfield  {author} {\bibinfo {author} {\bibfnamefont {M.~L.}\ \bibnamefont
  {Manning}}, \bibinfo {author} {\bibfnamefont {E.~G.}\ \bibnamefont {Daub}},
  \bibinfo {author} {\bibfnamefont {J.~S.}\ \bibnamefont {Langer}}, \ and\
  \bibinfo {author} {\bibfnamefont {J.~M.}\ \bibnamefont {Carlson}},\ }\href
  {\doibase 10.1103/PhysRevE.79.016110} {\bibfield  {journal} {\bibinfo
  {journal} {Phys. Rev. E}\ }\textbf {\bibinfo {volume} {79}},\ \bibinfo
  {pages} {016110} (\bibinfo {year} {2009})}\BibitemShut {NoStop}%
\bibitem [{\citenamefont {Karmakar}\ \emph {et~al.}(2010)\citenamefont
  {Karmakar}, \citenamefont {Lerner},\ and\ \citenamefont
  {Procaccia}}]{Karmakar2010}%
  \BibitemOpen
  \bibfield  {author} {\bibinfo {author} {\bibfnamefont {S.}~\bibnamefont
  {Karmakar}}, \bibinfo {author} {\bibfnamefont {E.}~\bibnamefont {Lerner}}, \
  and\ \bibinfo {author} {\bibfnamefont {I.}~\bibnamefont {Procaccia}},\ }\href
  {\doibase 10.1103/PhysRevE.82.055103} {\bibfield  {journal} {\bibinfo
  {journal} {Phys. Rev. E}\ }\textbf {\bibinfo {volume} {82}},\ \bibinfo
  {pages} {055103} (\bibinfo {year} {2010})}\BibitemShut {NoStop}%
\bibitem [{\citenamefont {Barrat}\ and\ \citenamefont
  {Lema\^{\i}tre}(2011)}]{Barrat2011}%
  \BibitemOpen
  \bibfield  {author} {\bibinfo {author} {\bibfnamefont {J.~L.}\ \bibnamefont
  {Barrat}}\ and\ \bibinfo {author} {\bibfnamefont {A.}~\bibnamefont
  {Lema\^{\i}tre}},\ }in\ \href {\doibase
  10.1093/acprof:oso/9780199691470.003.0008} {\emph {\bibinfo {booktitle}
  {Dynamical heterogeneities in glasses, colloids, and granular media}}},\
  \bibinfo {editor} {edited by\ \bibinfo {editor} {\bibfnamefont
  {L.}~\bibnamefont {Berthier}}, \bibinfo {editor} {\bibfnamefont
  {G.}~\bibnamefont {Biroli}}, \bibinfo {editor} {\bibfnamefont {J.~P.}\
  \bibnamefont {Bouchaud}}, \bibinfo {editor} {\bibfnamefont {L.}~\bibnamefont
  {Cipelletti}}, \ and\ \bibinfo {editor} {\bibfnamefont {W.~v.}\ \bibnamefont
  {Saarloos}}}\ (\bibinfo  {publisher} {Oxford Science Publications},\ \bibinfo
  {address} {Oxford},\ \bibinfo {year} {2011})\ Chap.~\bibinfo {chapter} {8},
  pp.\ \bibinfo {pages} {264--297}\BibitemShut {NoStop}%
\bibitem [{\citenamefont {Dasgupta}\ \emph {et~al.}(2012)\citenamefont
  {Dasgupta}, \citenamefont {Hentschel},\ and\ \citenamefont
  {Procaccia}}]{Dasgupta2012}%
  \BibitemOpen
  \bibfield  {author} {\bibinfo {author} {\bibfnamefont {R.}~\bibnamefont
  {Dasgupta}}, \bibinfo {author} {\bibfnamefont {H.~G.~E.}\ \bibnamefont
  {Hentschel}}, \ and\ \bibinfo {author} {\bibfnamefont {I.}~\bibnamefont
  {Procaccia}},\ }\href {\doibase 10.1103/PhysRevLett.109.255502} {\bibfield
  {journal} {\bibinfo  {journal} {Phys. Rev. Lett.}\ }\textbf {\bibinfo
  {volume} {109}},\ \bibinfo {pages} {255502} (\bibinfo {year}
  {2012})}\BibitemShut {NoStop}%
\bibitem [{\citenamefont {Keim}\ and\ \citenamefont
  {Arratia}(2013)}]{Keim2013}%
  \BibitemOpen
  \bibfield  {author} {\bibinfo {author} {\bibfnamefont {N.~C.}\ \bibnamefont
  {Keim}}\ and\ \bibinfo {author} {\bibfnamefont {P.~E.}\ \bibnamefont
  {Arratia}},\ }\href {\doibase 10.1039/c3sm51014j} {\bibfield  {journal}
  {\bibinfo  {journal} {Soft Matter}\ }\textbf {\bibinfo {volume} {9}},\
  \bibinfo {pages} {6222} (\bibinfo {year} {2013})}\BibitemShut {NoStop}%
\bibitem [{\citenamefont {Lin}\ \emph {et~al.}(2014)\citenamefont {Lin},
  \citenamefont {Lerner}, \citenamefont {Rosso},\ and\ \citenamefont
  {Wyart}}]{Lin2014}%
  \BibitemOpen
  \bibfield  {author} {\bibinfo {author} {\bibfnamefont {J.}~\bibnamefont
  {Lin}}, \bibinfo {author} {\bibfnamefont {E.}~\bibnamefont {Lerner}},
  \bibinfo {author} {\bibfnamefont {A.}~\bibnamefont {Rosso}}, \ and\ \bibinfo
  {author} {\bibfnamefont {M.}~\bibnamefont {Wyart}},\ }\href {\doibase
  10.1073/pnas.1406391111} {\ \textbf {\bibinfo {volume} {111}},\ \bibinfo
  {pages} {14382} (\bibinfo {year} {2014})}\BibitemShut {NoStop}%
\bibitem [{\citenamefont {E.~D.~Knowlton}\ and\ \citenamefont
  {Cipelletti}(2014)}]{Knowlton2014}%
  \BibitemOpen
  \bibfield  {author} {\bibinfo {author} {\bibfnamefont {D.~J.~P.}\
  \bibnamefont {E.~D.~Knowlton}}\ and\ \bibinfo {author} {\bibfnamefont
  {L.}~\bibnamefont {Cipelletti}},\ }\href {\doibase 10.1039/C4SM00531G}
  {\bibfield  {journal} {\bibinfo  {journal} {Soft Matter}\ }\textbf {\bibinfo
  {volume} {10}},\ \bibinfo {pages} {6931} (\bibinfo {year}
  {2014})}\BibitemShut {NoStop}%
\bibitem [{\citenamefont {Hima~Nagamanasa}\ \emph {et~al.}(2014)\citenamefont
  {Hima~Nagamanasa}, \citenamefont {Gokhale}, \citenamefont {Sood},\ and\
  \citenamefont {Ganapathy}}]{Nagamanasa2014}%
  \BibitemOpen
  \bibfield  {author} {\bibinfo {author} {\bibfnamefont {K.}~\bibnamefont
  {Hima~Nagamanasa}}, \bibinfo {author} {\bibfnamefont {S.}~\bibnamefont
  {Gokhale}}, \bibinfo {author} {\bibfnamefont {A.~K.}\ \bibnamefont {Sood}}, \
  and\ \bibinfo {author} {\bibfnamefont {R.}~\bibnamefont {Ganapathy}},\ }\href
  {\doibase 10.1103/PhysRevE.89.062308} {\bibfield  {journal} {\bibinfo
  {journal} {Phys. Rev. E}\ }\textbf {\bibinfo {volume} {89}},\ \bibinfo
  {pages} {062308} (\bibinfo {year} {2014})}\BibitemShut {NoStop}%
\bibitem [{\citenamefont {Denisov}\ \emph {et~al.}(2015)\citenamefont
  {Denisov}, \citenamefont {Dang}, \citenamefont {Struth}, \citenamefont
  {Zaccone}, \citenamefont {Wegdam},\ and\ \citenamefont
  {Schall}}]{denisov2015}%
  \BibitemOpen
  \bibfield  {author} {\bibinfo {author} {\bibfnamefont {D.~V.}\ \bibnamefont
  {Denisov}}, \bibinfo {author} {\bibfnamefont {M.~T.}\ \bibnamefont {Dang}},
  \bibinfo {author} {\bibfnamefont {B.}~\bibnamefont {Struth}}, \bibinfo
  {author} {\bibfnamefont {A.}~\bibnamefont {Zaccone}}, \bibinfo {author}
  {\bibfnamefont {G.~H.}\ \bibnamefont {Wegdam}}, \ and\ \bibinfo {author}
  {\bibfnamefont {P.}~\bibnamefont {Schall}},\ }\href@noop {} {\bibfield
  {journal} {\bibinfo  {journal} {Scientific reports}\ }\textbf {\bibinfo
  {volume} {5}} (\bibinfo {year} {2015})}\BibitemShut {NoStop}%
\bibitem [{\citenamefont {Regev}\ \emph {et~al.}(2015)\citenamefont {Regev},
  \citenamefont {Weber}, \citenamefont {Reichhardt}, \citenamefont {Dahmen},\
  and\ \citenamefont {Lookman}}]{regev2015reversibility}%
  \BibitemOpen
  \bibfield  {author} {\bibinfo {author} {\bibfnamefont {I.}~\bibnamefont
  {Regev}}, \bibinfo {author} {\bibfnamefont {J.}~\bibnamefont {Weber}},
  \bibinfo {author} {\bibfnamefont {C.}~\bibnamefont {Reichhardt}}, \bibinfo
  {author} {\bibfnamefont {K.~A.}\ \bibnamefont {Dahmen}}, \ and\ \bibinfo
  {author} {\bibfnamefont {T.}~\bibnamefont {Lookman}},\ }\href@noop {}
  {\bibfield  {journal} {\bibinfo  {journal} {Nature communications}\ }\textbf
  {\bibinfo {volume} {6}} (\bibinfo {year} {2015})}\BibitemShut {NoStop}%
\bibitem [{\citenamefont {Liu}\ \emph {et~al.}(2016)\citenamefont {Liu},
  \citenamefont {Ferrero}, \citenamefont {Puosi}, \citenamefont {Barrat},\ and\
  \citenamefont {Martens}}]{Liu2016}%
  \BibitemOpen
  \bibfield  {author} {\bibinfo {author} {\bibfnamefont {C.}~\bibnamefont
  {Liu}}, \bibinfo {author} {\bibfnamefont {E.~E.}\ \bibnamefont {Ferrero}},
  \bibinfo {author} {\bibfnamefont {F.}~\bibnamefont {Puosi}}, \bibinfo
  {author} {\bibfnamefont {J.-L.}\ \bibnamefont {Barrat}}, \ and\ \bibinfo
  {author} {\bibfnamefont {K.}~\bibnamefont {Martens}},\ }\href {\doibase
  10.1103/PhysRevLett.116.065501} {\bibfield  {journal} {\bibinfo  {journal}
  {Phys. Rev. Lett.}\ }\textbf {\bibinfo {volume} {116}},\ \bibinfo {pages}
  {065501} (\bibinfo {year} {2016})}\BibitemShut {NoStop}%
\bibitem [{\citenamefont {Jaiswal}\ \emph {et~al.}(2016)\citenamefont
  {Jaiswal}, \citenamefont {Procaccia}, \citenamefont {Rainone},\ and\
  \citenamefont {Singh}}]{Itamar2016yielding}%
  \BibitemOpen
  \bibfield  {author} {\bibinfo {author} {\bibfnamefont {P.~K.}\ \bibnamefont
  {Jaiswal}}, \bibinfo {author} {\bibfnamefont {I.}~\bibnamefont {Procaccia}},
  \bibinfo {author} {\bibfnamefont {C.}~\bibnamefont {Rainone}}, \ and\
  \bibinfo {author} {\bibfnamefont {M.}~\bibnamefont {Singh}},\ }\href
  {\doibase 10.1103/PhysRevLett.116.085501} {\bibfield  {journal} {\bibinfo
  {journal} {Phys. Rev. Lett.}\ }\textbf {\bibinfo {volume} {116}},\ \bibinfo
  {pages} {085501} (\bibinfo {year} {2016})}\BibitemShut {NoStop}%
\bibitem [{\citenamefont {Shrivastav}\ \emph {et~al.}(2016)\citenamefont
  {Shrivastav}, \citenamefont {Chaudhuri},\ and\ \citenamefont
  {Horbach}}]{shrivastav2016yielding}%
  \BibitemOpen
  \bibfield  {author} {\bibinfo {author} {\bibfnamefont {G.~P.}\ \bibnamefont
  {Shrivastav}}, \bibinfo {author} {\bibfnamefont {P.}~\bibnamefont
  {Chaudhuri}}, \ and\ \bibinfo {author} {\bibfnamefont {J.}~\bibnamefont
  {Horbach}},\ }\href {\doibase 10.1103/PhysRevE.94.042605} {\bibfield
  {journal} {\bibinfo  {journal} {Phys. Rev. E}\ }\textbf {\bibinfo {volume}
  {94}},\ \bibinfo {pages} {042605} (\bibinfo {year} {2016})}\BibitemShut
  {NoStop}%
\bibitem [{\citenamefont {Leishangthem}\ \emph {et~al.}(2017)\citenamefont
  {Leishangthem}, \citenamefont {Parmar},\ and\ \citenamefont
  {Sastry}}]{leishangthem2017yielding}%
  \BibitemOpen
  \bibfield  {author} {\bibinfo {author} {\bibfnamefont {P.}~\bibnamefont
  {Leishangthem}}, \bibinfo {author} {\bibfnamefont {A.~D.}\ \bibnamefont
  {Parmar}}, \ and\ \bibinfo {author} {\bibfnamefont {S.}~\bibnamefont
  {Sastry}},\ }\href@noop {} {\bibfield  {journal} {\bibinfo  {journal} {Nature
  communications}\ }\textbf {\bibinfo {volume} {8}},\ \bibinfo {pages} {14653}
  (\bibinfo {year} {2017})}\BibitemShut {NoStop}%
\bibitem [{\citenamefont {Parisi}\ \emph {et~al.}(2017)\citenamefont {Parisi},
  \citenamefont {Procaccia}, \citenamefont {Rainone},\ and\ \citenamefont
  {Singh}}]{parisi2017shear}%
  \BibitemOpen
  \bibfield  {author} {\bibinfo {author} {\bibfnamefont {G.}~\bibnamefont
  {Parisi}}, \bibinfo {author} {\bibfnamefont {I.}~\bibnamefont {Procaccia}},
  \bibinfo {author} {\bibfnamefont {C.}~\bibnamefont {Rainone}}, \ and\
  \bibinfo {author} {\bibfnamefont {M.}~\bibnamefont {Singh}},\ }\href
  {\doibase 10.1073/pnas.1700075114} {\bibfield  {journal} {\bibinfo  {journal}
  {Proceedings of the National Academy of Sciences}\ }\textbf {\bibinfo
  {volume} {114}},\ \bibinfo {pages} {5577} (\bibinfo {year}
  {2017})}\BibitemShut {NoStop}%
\bibitem [{\citenamefont {Urbani}\ and\ \citenamefont
  {Zamponi}(2017)}]{Urbani2017b}%
  \BibitemOpen
  \bibfield  {author} {\bibinfo {author} {\bibfnamefont {P.}~\bibnamefont
  {Urbani}}\ and\ \bibinfo {author} {\bibfnamefont {F.}~\bibnamefont
  {Zamponi}},\ }\href@noop {} {\bibfield  {journal} {\bibinfo  {journal} {Phys.
  Rev. Lett.}\ }\textbf {\bibinfo {volume} {118}},\ \bibinfo {pages} {038001}
  (\bibinfo {year} {2017})}\BibitemShut {NoStop}%
\bibitem [{\citenamefont {Jin}\ \emph {et~al.}(2018)\citenamefont {Jin},
  \citenamefont {Urbani}, \citenamefont {Zamponi},\ and\ \citenamefont
  {Yoshino}}]{Jin}%
  \BibitemOpen
  \bibfield  {author} {\bibinfo {author} {\bibfnamefont {Y.}~\bibnamefont
  {Jin}}, \bibinfo {author} {\bibfnamefont {P.}~\bibnamefont {Urbani}},
  \bibinfo {author} {\bibfnamefont {F.}~\bibnamefont {Zamponi}}, \ and\
  \bibinfo {author} {\bibfnamefont {H.}~\bibnamefont {Yoshino}},\ }\href@noop
  {} {\bibfield  {journal} {\bibinfo  {journal} {arXiv preprint
  arXiv:1803.04597}\ } (\bibinfo {year} {2018})}\BibitemShut {NoStop}%
\bibitem [{\citenamefont {Ozawa}\ \emph {et~al.}(2018)\citenamefont {Ozawa},
  \citenamefont {Berthier}, \citenamefont {Biroli}, \citenamefont {Rosso},\
  and\ \citenamefont {Tarjus}}]{Ozawa2018a}%
  \BibitemOpen
  \bibfield  {author} {\bibinfo {author} {\bibfnamefont {M.}~\bibnamefont
  {Ozawa}}, \bibinfo {author} {\bibfnamefont {L.}~\bibnamefont {Berthier}},
  \bibinfo {author} {\bibfnamefont {G.}~\bibnamefont {Biroli}}, \bibinfo
  {author} {\bibfnamefont {A.}~\bibnamefont {Rosso}}, \ and\ \bibinfo {author}
  {\bibfnamefont {G.}~\bibnamefont {Tarjus}},\ }\href@noop {} {\bibfield
  {journal} {\bibinfo  {journal} {arXiv preprint arXiv:1803.11502}\ } (\bibinfo
  {year} {2018})}\BibitemShut {NoStop}%
\bibitem [{\citenamefont {Popovi{\'c}}\ \emph {et~al.}(2018)\citenamefont
  {Popovi{\'c}}, \citenamefont {de~Geus},\ and\ \citenamefont
  {Wyart}}]{Popovic2018a}%
  \BibitemOpen
  \bibfield  {author} {\bibinfo {author} {\bibfnamefont {M.}~\bibnamefont
  {Popovi{\'c}}}, \bibinfo {author} {\bibfnamefont {T.~W.}\ \bibnamefont
  {de~Geus}}, \ and\ \bibinfo {author} {\bibfnamefont {M.}~\bibnamefont
  {Wyart}},\ }\href@noop {} {\bibfield  {journal} {\bibinfo  {journal} {arXiv
  preprint arXiv:1803.11504}\ } (\bibinfo {year} {2018})}\BibitemShut {NoStop}%
\bibitem [{\citenamefont {Argon}(1979)}]{argon1979plastic}%
  \BibitemOpen
  \bibfield  {author} {\bibinfo {author} {\bibfnamefont {A.}~\bibnamefont
  {Argon}},\ }\href {\doibase https://doi.org/10.1016/0001-6160(79)90055-5}
  {\bibfield  {journal} {\bibinfo  {journal} {Acta metallurgica}\ }\textbf
  {\bibinfo {volume} {27}},\ \bibinfo {pages} {47} (\bibinfo {year}
  {1979})}\BibitemShut {NoStop}%
\bibitem [{\citenamefont {Falk}\ and\ \citenamefont
  {Langer}(1998)}]{falk1998dynamics}%
  \BibitemOpen
  \bibfield  {author} {\bibinfo {author} {\bibfnamefont {M.~L.}\ \bibnamefont
  {Falk}}\ and\ \bibinfo {author} {\bibfnamefont {J.~S.}\ \bibnamefont
  {Langer}},\ }\href {\doibase 10.1103/PhysRevE.57.7192} {\bibfield  {journal}
  {\bibinfo  {journal} {Phys. Rev. E}\ }\textbf {\bibinfo {volume} {57}},\
  \bibinfo {pages} {7192} (\bibinfo {year} {1998})}\BibitemShut {NoStop}%
\bibitem [{\citenamefont {Picard}\ \emph {et~al.}(2004)\citenamefont {Picard},
  \citenamefont {Ajdari}, \citenamefont {Lequeux},\ and\ \citenamefont
  {Bocquet}}]{Picard2004}%
  \BibitemOpen
  \bibfield  {author} {\bibinfo {author} {\bibfnamefont {G.}~\bibnamefont
  {Picard}}, \bibinfo {author} {\bibfnamefont {A.}~\bibnamefont {Ajdari}},
  \bibinfo {author} {\bibfnamefont {F.}~\bibnamefont {Lequeux}}, \ and\
  \bibinfo {author} {\bibfnamefont {L.}~\bibnamefont {Bocquet}},\ }\href
  {\doibase 10.1140/epje/i2004-10054-8} {\bibfield  {journal} {\bibinfo
  {journal} {The European Physical Journal E}\ }\textbf {\bibinfo {volume}
  {15}},\ \bibinfo {pages} {371} (\bibinfo {year} {2004})}\BibitemShut
  {NoStop}%
\bibitem [{\citenamefont {Puosi}\ \emph {et~al.}(2014)\citenamefont {Puosi},
  \citenamefont {Rottler},\ and\ \citenamefont {Barrat}}]{Puosi2014a}%
  \BibitemOpen
  \bibfield  {author} {\bibinfo {author} {\bibfnamefont {F.}~\bibnamefont
  {Puosi}}, \bibinfo {author} {\bibfnamefont {J.}~\bibnamefont {Rottler}}, \
  and\ \bibinfo {author} {\bibfnamefont {J.-L.}\ \bibnamefont {Barrat}},\
  }\href@noop {} {\bibfield  {journal} {\bibinfo  {journal} {Phys. Rev. E}\
  }\textbf {\bibinfo {volume} {89}},\ \bibinfo {pages} {042302} (\bibinfo
  {year} {2014})}\BibitemShut {NoStop}%
\bibitem [{\citenamefont {Talamali}\ \emph {et~al.}(2012)\citenamefont
  {Talamali}, \citenamefont {Pet{\"a}j{\"a}}, \citenamefont {Vandembroucq},\
  and\ \citenamefont {Roux}}]{Talamali2012c}%
  \BibitemOpen
  \bibfield  {author} {\bibinfo {author} {\bibfnamefont {M.}~\bibnamefont
  {Talamali}}, \bibinfo {author} {\bibfnamefont {V.}~\bibnamefont
  {Pet{\"a}j{\"a}}}, \bibinfo {author} {\bibfnamefont {D.}~\bibnamefont
  {Vandembroucq}}, \ and\ \bibinfo {author} {\bibfnamefont {S.}~\bibnamefont
  {Roux}},\ }\href@noop {} {\bibfield  {journal} {\bibinfo  {journal} {Comptes
  Rendus Mecanique}\ }\textbf {\bibinfo {volume} {340}},\ \bibinfo {pages}
  {275} (\bibinfo {year} {2012})}\BibitemShut {NoStop}%
\bibitem [{\citenamefont {Tyukodi}\ \emph {et~al.}(2016)\citenamefont
  {Tyukodi}, \citenamefont {Patinet}, \citenamefont {Roux},\ and\ \citenamefont
  {Vandembroucq}}]{Tyukodi2016c}%
  \BibitemOpen
  \bibfield  {author} {\bibinfo {author} {\bibfnamefont {B.}~\bibnamefont
  {Tyukodi}}, \bibinfo {author} {\bibfnamefont {S.}~\bibnamefont {Patinet}},
  \bibinfo {author} {\bibfnamefont {S.}~\bibnamefont {Roux}}, \ and\ \bibinfo
  {author} {\bibfnamefont {D.}~\bibnamefont {Vandembroucq}},\ }\href@noop {}
  {\bibfield  {journal} {\bibinfo  {journal} {Phys. Rev. E}\ }\textbf {\bibinfo
  {volume} {93}},\ \bibinfo {pages} {063005} (\bibinfo {year}
  {2016})}\BibitemShut {NoStop}%
\bibitem [{\citenamefont {Vasisht}\ \emph {et~al.}(2017)\citenamefont
  {Vasisht}, \citenamefont {Roberts},\ and\ \citenamefont
  {Del~Gado}}]{vasisht2017emergence}%
  \BibitemOpen
  \bibfield  {author} {\bibinfo {author} {\bibfnamefont {V.~V.}\ \bibnamefont
  {Vasisht}}, \bibinfo {author} {\bibfnamefont {G.}~\bibnamefont {Roberts}}, \
  and\ \bibinfo {author} {\bibfnamefont {E.}~\bibnamefont {Del~Gado}},\
  }\href@noop {} {\bibfield  {journal} {\bibinfo  {journal} {arXiv preprint
  arXiv:1709.08717}\ } (\bibinfo {year} {2017})}\BibitemShut {NoStop}%
\bibitem [{\citenamefont {Fiocco}\ \emph {et~al.}(2013)\citenamefont {Fiocco},
  \citenamefont {Foffi},\ and\ \citenamefont {Sastry}}]{Fiocco2013}%
  \BibitemOpen
  \bibfield  {author} {\bibinfo {author} {\bibfnamefont {D.}~\bibnamefont
  {Fiocco}}, \bibinfo {author} {\bibfnamefont {G.}~\bibnamefont {Foffi}}, \
  and\ \bibinfo {author} {\bibfnamefont {S.}~\bibnamefont {Sastry}},\
  }\href@noop {} {\bibfield  {journal} {\bibinfo  {journal} {Phys. Rev. E}\
  }\textbf {\bibinfo {volume} {88}},\ \bibinfo {pages} {020301} (\bibinfo
  {year} {2013})}\BibitemShut {NoStop}%
\bibitem [{\citenamefont {Fiocco}\ \emph {et~al.}(2014)\citenamefont {Fiocco},
  \citenamefont {Foffi},\ and\ \citenamefont {Sastry}}]{Fiocco2014}%
  \BibitemOpen
  \bibfield  {author} {\bibinfo {author} {\bibfnamefont {D.}~\bibnamefont
  {Fiocco}}, \bibinfo {author} {\bibfnamefont {G.}~\bibnamefont {Foffi}}, \
  and\ \bibinfo {author} {\bibfnamefont {S.}~\bibnamefont {Sastry}},\
  }\href@noop {} {\bibfield  {journal} {\bibinfo  {journal} {Phys. Rev. Lett.}\
  }\textbf {\bibinfo {volume} {112}},\ \bibinfo {pages} {025702} (\bibinfo
  {year} {2014})}\BibitemShut {NoStop}%
\bibitem [{\citenamefont {Radhakrishnan}\ and\ \citenamefont
  {Fielding}(2016)}]{Radhakrishnan2016b}%
  \BibitemOpen
  \bibfield  {author} {\bibinfo {author} {\bibfnamefont {R.}~\bibnamefont
  {Radhakrishnan}}\ and\ \bibinfo {author} {\bibfnamefont {S.~M.}\ \bibnamefont
  {Fielding}},\ }\href@noop {} {\bibfield  {journal} {\bibinfo  {journal}
  {Phys. Rev. Lett.}\ }\textbf {\bibinfo {volume} {117}},\ \bibinfo {pages}
  {188001} (\bibinfo {year} {2016})}\BibitemShut {NoStop}%
\bibitem [{\citenamefont {Kawasaki}\ and\ \citenamefont
  {Berthier}(2016)}]{kawasaki2016macroscopic}%
  \BibitemOpen
  \bibfield  {author} {\bibinfo {author} {\bibfnamefont {T.}~\bibnamefont
  {Kawasaki}}\ and\ \bibinfo {author} {\bibfnamefont {L.}~\bibnamefont
  {Berthier}},\ }\href {\doibase 10.1103/PhysRevE.94.022615} {\bibfield
  {journal} {\bibinfo  {journal} {Phys. Rev. E}\ }\textbf {\bibinfo {volume}
  {94}},\ \bibinfo {pages} {022615} (\bibinfo {year} {2016})}\BibitemShut
  {NoStop}%
\bibitem [{\citenamefont {Priezjev}(2018)}]{PRIEZJEV2018}%
  \BibitemOpen
  \bibfield  {author} {\bibinfo {author} {\bibfnamefont {N.~V.}\ \bibnamefont
  {Priezjev}},\ }\href {\doibase
  https://doi.org/10.1016/j.commatsci.2018.03.062} {\bibfield  {journal}
  {\bibinfo  {journal} {Computational Materials Science}\ }\textbf {\bibinfo
  {volume} {150}},\ \bibinfo {pages} {162 } (\bibinfo {year}
  {2018})}\BibitemShut {NoStop}%
\bibitem [{\citenamefont {Plimpton}(1995)}]{Plimpton1995}%
  \BibitemOpen
  \bibfield  {author} {\bibinfo {author} {\bibfnamefont {S.}~\bibnamefont
  {Plimpton}},\ }\href {\doibase https://doi.org/10.1006/jcph.1995.1039}
  {\bibfield  {journal} {\bibinfo  {journal} {Journal of Computational
  Physics}\ }\textbf {\bibinfo {volume} {117}},\ \bibinfo {pages} {1 }
  (\bibinfo {year} {1995})}\BibitemShut {NoStop}%
\bibitem [{\citenamefont {Lacks}\ and\ \citenamefont
  {Osborne}(2004)}]{Lacks2004a}%
  \BibitemOpen
  \bibfield  {author} {\bibinfo {author} {\bibfnamefont {D.~J.}\ \bibnamefont
  {Lacks}}\ and\ \bibinfo {author} {\bibfnamefont {M.~J.}\ \bibnamefont
  {Osborne}},\ }\href {\doibase 10.1103/PhysRevLett.93.255501} {\bibfield
  {journal} {\bibinfo  {journal} {Phys. Rev. Lett.}\ }\textbf {\bibinfo
  {volume} {93}},\ \bibinfo {pages} {1} (\bibinfo {year} {2004})}\BibitemShut
  {NoStop}%
\end{thebibliography}%

\end{document}